\documentclass[final,3p,times,12pt]{elsarticle}
\usepackage{amsmath,amssymb}
\usepackage{graphicx}
\usepackage{rotating}
\usepackage{color}
\usepackage{xcolor}
\usepackage{fontenc}
\usepackage{slashed}
\usepackage{longtable}
\usepackage{dcolumn}
\usepackage{bm}
\usepackage{natbib}
\usepackage{lineno,hyperref}
\hypersetup{colorlinks,bookmarksopen,bookmarksnumbered,linkcolor=blue,pdfstartview=FitH,urlcolor=blue,citecolor=blue}

\journal{arXiv}

\begin{document}

\begin{frontmatter}
\title{Restricting Majorana phases in the Tri-Bi-Maximal limit}

\author[1]{Diana C. Rivera-Agudelo\corref{cor1}}
\cortext[cor1]{Corresponding author.}
\ead{diana.rivera11@usc.edu.co}
\address[1]{Universidad Santiago de Cali, Facultad de Ciencias B\'asicas, Campus Pampalinda, Calle 5 No. 62-00, C\'odigo Postal 76001, Santiago de Cali, Colombia}

\author[1,2]{S. L. Tostado}
\ead{sergio.tostado@correounivalle.edu.co}
\address[2]{Departamento de F\'isica, Universidad del Valle, Cll. 13 \# 100 - 0 A.A. 24360, Santiago de Cali, Colombia}

\begin{abstract}
 The Tri-Bi-Maximal pattern has been long investigated as the symmetric scenario that lies behind the neutrino mixing matrix. It predicts a null reactor angle and hence forbids $CP$ violation in the lepton sector, which is in contrast to the current experimental determinations. We explore different deviations from this pattern to restore the compatibility with the latest fits of neutrino mixing parameters. We consider two unitary matrices to correct the symmetric pattern, each of them is written in terms of one single angle and one complex phase, which will be constrained by the experimental mixings and from symmetry restrictions in the mass matrix. We note that these correction parameters would allow us to obtain simultaneous information about the Dirac and Majorana $CP$ phases in some specific scenarios. We show that the predicted values lead to sharped regions for the neutrinoless double beta decay amplitude, in the selected cases, that could be tested with forthcoming results.  
\end{abstract}

\end{frontmatter}

%\maketitle

\section{Introduction}\label{sec:intro} 
 Neutrino oscillation phenomena are described by the Pontecorvo-Maki-Nakagawa-Sakata (PMNS) matrix \cite{Pontecorvo:1957cp,Maki:1962mu} which relates neutrino flavor eigenstates with mass eigenstates. The PMNS matrix is written in the standard form in terms of three mixing angles and three $CP$-violating ($CPV$) phases, in the case of Majorana neutrinos. Neutrino mixing angles have been determined from experimental results with a precision of a few percent \cite{Tanabashi:2018oca,Capozzi:2018ubv,deSalas:2017kay,Esteban:2018azc}, where some hits for a Dirac $CP$ phase ($\delta_{CP}$) different from zero have also appeared. It is expected that future neutrino experiments could help to confirm this expectation \cite{Abi:2018dnh}. Nevertheless, accessing Majorana phases ($\beta_1$ and $\beta_2$) is currently a difficult task, where, future improvements in neutrinoless double beta decay experiments may shed some light in this respect \cite{Dolinski:2019nrj}. Neutrino mixing matrix is usually parametrized as 
\begin{eqnarray}
\label{PMNS}
U_{\rm PMNS} &=&  \left( \begin{array}{ccc} 
c_{12} c_{13} & s_{12}c_{13} & s_{13} e^{-i\delta_{CP}} \\
- s_{12} c_{23} + c_{12}s_{23}s_{13}e^{i\delta_{CP}}   & 
c_{12}c_{23} + s_{12}s_{23}s_{13}e^{i\delta_{CP}} & -s_{23}c_{13} \\
-s_{12}s_{23} - c_{12}c_{23}s_{13}e^{i\delta_{CP}} & 
c_{12}s_{23} -c_{23}s_{12}s_{13}e^{i\delta_{CP}} & c_{23}c_{13} 
\end{array} \right) \nonumber \\ 
&&\times {\rm diag} \left[ 1,  e^{-i\frac{\beta_1}{2}}, e^{-i\frac{\beta_2}{2}}\right]~,
\end{eqnarray}
where, $c_{ij}$ and $s_{ij}$ stand for $\cos \theta_{ij}$ and $\sin \theta_{ij}$, respectively. $\theta_{ij}$ refers to the mixing angles $\theta_{12}$, $\theta_{13}$, and $\theta_{23}$, whereas $\delta_{CP}$, $\beta_1$ and $\beta_2$ denote the $CPV$ phases. 

Historically, the very first results on neutrino oscillations seemed to be consistent with $\theta_{13} =0$ and $\theta_{23}=\pi/4$. Such a particular selection allows the PMNS matrix in Eq. (\ref{PMNS}) to take the form 
\begin{equation}\label{Umutau}
U_{\mu-\tau} =  \left( \begin{array}{ccc} 
c_{12}   & s_{12}   & 0 \\
\frac{-s_{12}}{\sqrt{2}} & \frac{c_{12}}{\sqrt{2}} & \frac{- 1}{\sqrt{2}} \\
\frac{- s_{12}}{\sqrt{2}} & \frac{ c_{12}}{\sqrt{2}}& \frac{1}{\sqrt{2}}
\end{array} \right),
\end{equation}
which reflects the relation $|U_{\mu i}| = |U_{\tau i}|$ between the second and third row, also known as a $\mu-\tau$ interchange symmetry. Such a specific form of the mixing matrix has inspired a large number of theoretical works based on this symmetry pattern \cite{Fukuyama:1997ky,Harrison:2002et,Babu:2002ex,Ohlsson:2002rb,Mohapatra:2004mf,Ghosal:2004qb,Ma:2004zv,Mohapatra:2006un,Joshipura:2005vy,Fuki:2006xw,Riazuddin:2007aa,Luhn:2007sy,Koide:2008zza,Honda:2008rs,Ishimori:2008gp,Ge:2010js,Ge:2011ih,He:2011kn,Ge:2011qn,Talbert:2014bda,He:2012yt,Stancu:1999ct}. Among the different posibilities in fixing the remaining parameter, the well known Tri-Bi-Maximal (TBM) pattern, where  $\sin^2 \theta_{12}=1/3$, was also proposed \cite{Vissani:1997pa} according to the experimental evidence. The posbibility of generating this specific pattern from larger flavor symmetries is also of great theoretical interest and has been investigated previously \cite{Altarelli:2012ss,Xing:2015fdg}. Nevertheless, as it is direct to note, such a symmetric pattern is already excluded by most recent global fits of mixing parameters \cite{Tanabashi:2018oca,Capozzi:2018ubv,deSalas:2017kay,Esteban:2018azc}.

 Concerning the $\mu-\tau$ symmetry, some deviations from the TBM pattern have been investigated in order to fit the experimental results \cite{Petcov:2014laa,Garg:2013xwa,Garg:2017mjk,Shimizu:2014ria,Sruthilaya2015,Sruthilaya2016,Rahat:2018sgs}. Parametrizations of the form 
\begin{equation}\label{eq:corr}
U_{\rm PMNS} = U_{TBM}U_{\rm Corr}
\end{equation}
have been studied, with different proposals for the correction matrix $U_{\rm Corr}$. In Eq. \ref{eq:corr}, $U_{TBM}$ denotes the fully symmetric TBM matrix while $U_{\rm Corr}$ encodes the deviations from the symmetric pattern that can be parameterized in different ways. For instance, it has been shown in Refs. \cite{Garg:2013xwa,Garg:2017mjk} that corrections in terms of one or two orthogonal rotations can predict mixing angles within the experimental range ($3\sigma$). In addition, a unitary correction matrix can also give important information about the Dirac $CP$ phase \cite{Shimizu:2014ria,Sruthilaya2015,Sruthilaya2016}. On the other hand, the possibility of analyzing the Majorana case have been explored in \cite{Petcov:2014laa,Chen:2018eou,Chen:2018zbq, Rivera-Agudelo:2019seg} by including additional phases in the corrected matrix. 

 It has been pointed out that a correction matrix of the form in Eq. (\ref{eq:corr}) does not necessarily lead to small deviations from the $\mu-\tau$ symmetry in the mass matrix even for small correction angles \cite{Rivera-Agudelo:2019seg}. As noticed in \cite{Gupta:2013it,Samanta:2018hqm,Rivera2}, restricting the corrections parameters to small deviations from the symmetric mass matrix may give additional constraints on the Majorana phases. Considering this additional restriction could help to bound the correction parameters in $U_{\rm Corr}$ and hence modify the predicted mixing parameters.

The paper is organized as follows. In Section \ref{sec:deviations} we investigate the different parametrizations which correct the TBM mixing pattern. For the cases of interest, some analytical relations are presented for the mixings parameters in terms of the correction parameters. In Section \ref{sec:mutau} we define an approximate $\mu-\tau$ symmetry in the mass matrix in terms of two breaking parameters. In this respect, we also study the effects of corrections angles over the neutrino mass matrix. We present in Section \ref{sec:results} our main numerical estimations of $CPV$ phases and their phenomenological implications on the neutrinoless double beta decay amplitude ($|m_{ee}|$). Finally, in Section \ref{sec:conclusions}, we give our final comments and conclusions.

\section{Corrections to the TBM pattern}\label{sec:deviations}

 The recent estimations of neutrino parameters are in tension with the predictions of the TBM mixing pattern obtained from the $\mu-\tau$ symmetric matrix in Eq. (\ref{Umutau}) since it predicts $\theta_{13} = 0$ and hence $CP$ conservation. In this section, we introduce a particular form of departing from the TBM matrix motivated by the neediness to include $CP$ violating phases in our description.

Let us consider in the forthcoming analysis only deviations to the TBM pattern coming from the neutrino sector in the basis where the charged leptons are diagonal. To include $CP$ violation effects in our description, we should expect that such corrections can be written in terms of a complex matrix. Let us propose the following parametrization for the correction matrix in Eq. (\ref{eq:corr}) 
\begin{equation}\label{Ucorr}
U_{\rm Corr} = U_{ij}(\phi,\sigma) U_{kl}(\phi',\sigma') ~,
\end{equation}
where $U_{ij}$ ($U_{kl}$) is a unitary matrix which depends on one rotation angle $\phi$ ($\phi'$) and one complex phase $\sigma$ ($\sigma'$). Here $i,j,k,l=1,2,3$, $i\neq j$ and $k\neq l$. Eq. (\ref{Ucorr}) incorporates two new complex phases, which are aimed to be related to the physical Dirac and Majorana $CP$ phases. The rotation angles are expected to modulate the deviation from the TBM pattern and restore, if possible, the compatibility with the experimental angles. 
%Such type of parametrizations could be generated in a model dependent way from larger specific flavor symmetries \cite{Petcov:2014laa}.

The matrix in Eq. (\ref{Ucorr}) can be written in six different ways by considering combinations of the following rotation matrices 
\begin{equation}\label{rot13}
U_{13} = \left( \begin{array}{ccc} 
\cos \phi & 0 & \sin \phi ~e^{-i\sigma} \\
0   & 1 & 0 \\
-\sin \phi ~e^{i\sigma} & 0 & \cos \phi \end{array} \right),
\end{equation}
\begin{equation}\label{rot12}
U_{12} = \left( \begin{array}{ccc} 
\cos \phi & \sin \phi ~e^{-i\sigma} & 0 \\
-\sin \phi ~e^{i\sigma} & \cos \phi & 0 \\
0 & 0 & 1 \end{array} \right),
\end{equation}
and
\begin{equation}\label{rot23}
U_{23} = \left( \begin{array}{ccc} 
1 & 0 & 0 \\
0   & \cos \phi & \sin \phi ~e^{-i\sigma} \\
0 & -\sin \phi ~e^{i\sigma} & \cos \phi \end{array} \right) .
\end{equation}
In what follows, it is convenient to notice that the angle $\phi$, and the phase $\sigma$, should be renamed according to the conventions in Eq. (\ref{Ucorr}). In addition, it is straightforward to show that for combinations of the form $U_{ij}(0,0) U_{kl}(\phi',\sigma')$ and $U_{ij}(\phi,\sigma) U_{kl}(0,0)$, we recover the case of one unitary correction matrix that has been previously studied in \cite{Shimizu:2014ria,Sruthilaya2015,Sruthilaya2016,Rivera-Agudelo:2019seg} and will not be considered here. The trivial case $U_{ij}(0,0) U_{kl}(0,0)$ leaves the correction matrix as the identity matrix and the TBM pattern is then recovered.

The connection between the experimental angles in Eq. (\ref{PMNS}) and the correction parameters can be obtained from the elements of the corrected TBM matrix (Eq. \ref{eq:corr}) through the following relations%
\begin{equation}\label{mixings}
\sin^2{\theta_{12}}=\frac{\mid U_{e2} \mid^2}{1-\mid U_{e3} \mid^2}, ~~~~~\sin^2{\theta_{23}}=\frac{\mid U_{\mu 3} \mid^2}{1-\mid U_{e3} \mid^2}, ~~~~~\sin^2{\theta_{13}}=\mid U_{e3} \mid^2 .
\end{equation}
On the other hand, from the Jarlskog invariant ($J_{CP}$), we can relate the corrected matrix elements with the Dirac $CP$ phase $\delta_{CP}$ by	
\begin{eqnarray}\label{Jarskog}
J_{CP} &=& {\rm Im} \left[U_{e1} U_{\mu 2} U_{e2}^{*} U_{\mu 1}^{*} \right]\\
&=&(1-s^2{\theta_{13}})\sqrt{s^2{\theta_{13}}s^2{\theta_{12}}s^2{\theta_{23}}(1-s^2{\theta_{12}})(1-s^2{\theta_{23}})}  \sin\delta_{CP}.  \nonumber 
\end{eqnarray}
In addition, from the remaining invariants 
\begin{eqnarray}\label{invariants}
I_1 &=& {\rm Im}\left[ U_{e2}^2 U_{e1}^{*2} \right]= -\cos^2 \theta_{12} \cos^{4}\theta_{13} \sin^{2} \theta_{12} \sin \beta_1 \nonumber \\
I_2 &=& {\rm Im}\left[ U_{e3}^2 U_{e1}^{*2} \right]= -\cos^2 \theta_{12} \cos^2 \theta_{13} \sin^2 \theta_{13} \sin(\beta_2 + 2\delta_{CP}) ~,
\end{eqnarray}
we can, besides, to link together the Majorana $CP$ phases with the parameters in Eq. (\ref{eq:corr}) and hence with the mixing angles. 

From Eqs. (\ref{mixings}), (\ref{Jarskog}) and (\ref{invariants}), we should expect that the current fits of the mixing angles give the major restriction over the correction parameters ($\phi^{(\prime)},~ \sigma^{(\prime)}$), and that such allowed combinations, in the most optimistic scenario, lead to well-defined regions for the $CP$ phases. The predictions for the Dirac phase can be directly compared with its most recent global fit, but in the case of Majorana phases, we can only explore their combined ($\beta_1$ and $\beta_2$) effects through the neutrinoless double beta decay amplitude.

\section{$\mu - \tau$ symmetry  in the mass matrix}\label{sec:mutau}

Going back to the $\mu-\tau$ symmetric scheme of the mixing matrix (Eq. \ref{Umutau}), we can observe that the neutrino mass matrix obtained from $M_{\nu}=U_{\mu-\tau}{\rm diag}(m_1,m_2,m_3) U_{\mu-\tau}^{\rm T}$ reflects an exchange symmetry between the $\mu$ and $\tau$ entries, {\it i. e.} $|m_{e\mu}|=| m_{e \tau}|$ and $m_{\mu \mu}=m_{\tau \tau}$, which define the structure of a $\mu-\tau$ symmetric mass matrix\footnote{These relations are obtained for any $\mu-\tau$ symmetric mass matrix, including the TBM limit.}. It is worth noting that any change in the form of the mixing matrix, as in Eq. (\ref{eq:corr}), will change, in consequence, the symmetric structure of the neutrino mass matrix. As it has previously discussed \cite{Gupta:2013it,Rivera1}, departures from the $\mu-\tau$ symmetric mass matrix can be encoded in two breaking parameters. Therefore, we should be able to write down the neutrino mass matrix corresponding to the corrected mixing matrix in the form
\begin{equation}
M_\nu = M_{\mu - \tau} + \delta M \left(\hat{\delta},\hat \epsilon \right).
\end{equation}
Here, the matrix $M_{\mu-\tau}$ denotes the $\mu-\tau$ symmetric matrix, whereas $\delta M$ represent the deviations from the symmetric patter written in terms of two breaking parameters, $\hat{\delta}$ and $\hat \epsilon$. In the same way as the experimental angles, the breaking parameters can also be obtained from the matrix elements of Eq. (\ref{eq:corr}), namely \cite{Rivera1} 
\begin{eqnarray}\label{deltaepsilon}
\hat \delta &=& 
\frac{ \sum_i (U_{ei} U_{\tau i}-  U_{ei} U_{\mu i} ) m_i}{\sum_i U_{ei} U_{\mu 
i} m_i }~, \nonumber \\
\hat \epsilon &=&
\frac{\sum_i (U_{\tau i} U_{\tau i}-  
U_{\mu i} U_{\mu i}) m_i}{\sum_i U_{\mu i} U_{\mu i} m_i}~.
\end{eqnarray}
Following the approach in Ref. \cite{Gupta:2013it}, an approximate $\mu-\tau$ symmetric mass matrix is defined if we asked for $|\hat \delta| , |\hat \epsilon| \ll 1$. Therefore, in our approach, the requirement of an approximate symmetry in the mass matrix is then motivated by the necessity of implementing additional restrictions over the parameters modulating the departures from the TBM mixing and, in consequence, to reduce the predicted regions of the $CP$ phases.    

It is straightforward to show that the breaking parameters of Eq. (\ref{deltaepsilon}) can be written in terms of the correction parameters ($\phi^{(\prime)},~\sigma^{(\prime)}$) by using Eqs. (\ref{eq:corr}) and (\ref{Ucorr}). Within the different corrections, we should expect that the breaking parameters depend on the combination adopted. Although the complete expressions could be rather cumbersome, they can be directly obtained from Eq. (\ref{deltaepsilon}) for every parametrization. In addition, it is worth noting that the breaking parameters will also depend on the chosen neutrino mass hierarchy. Given that the absolute masses $|m_{1,2,3}|$ can be expressed in terms of the lightest neutrino mass $m_0$, we obtain
\begin{eqnarray}
|m_{2}| &=& \sqrt{m_0^2 + \Delta m_{21}^2} ~~,~~ |m_3| = \sqrt{m_0^2 + |\Delta m_{31}^2|}~~ {\rm for~ NH}, \nonumber \\
|m_{1}| &=& \sqrt{m_0^2 + |\Delta m_{31}^2|} ~~,~~ |m_2| = \sqrt{m_0^2 + |\Delta m_{31}^2| + \Delta m_{21}^2}~~ {\rm for~ IH},
\end{eqnarray}
where, $\Delta m_{21}^2 = m_2^2-m_1^2$ is usually known as the solar mass scale, and $\Delta m_{31}^2=m_3^2 - m_1^2$ is the atmospheric mass scale. The lightest neutrino mass, $m_0$, becomes $|m_1|$ for the normal mass hierarchy (NH), and $m_3$ for the inverted mass hierarchy (IH). In consequence, the expressions in Eq. (\ref{deltaepsilon}) could provide a direct connection between the correction parameters of the mixing matrix and the breaking parameters of the mass matrix.

%%%%%%%%%%%%%%%%%%%%%%%%%%%%%%%%%%%%%%%%%%
\section{Numerical Analysis}\label{sec:results}
%%%%%%%%%%%%%%%%%%%%%%%%%%%%%%%%%%%%%%%%%%
In this section, we present the complete analysis of the different combinations of unitary rotations for the corrected TBM matrix. For our numerical inputs, we use the results of the latest global fit of the various neutrino oscillation experiments from \cite{deSalas:2017kay}. In the analysis, we will consider the $3\sigma$ intervals of neutrino mixing angles and the squared mass differences of both mass hierarchies. 

%The experimetal intervals of the mixing angles, up to $3\sigma$, read
%\begin{eqnarray}\label{rangesIH}
%0.273<&\sin^2 \theta_{12}&<0.379~~, \nonumber \\
%0.0199<&\sin^2 \theta_{13}& <0.0244~~, \\
%0.453<&\sin^2 \theta_{23}&<0.598 ~~, \nonumber
%\end{eqnarray}
%while the squared mass differences are $\Delta m_{21}^2 = 7.55^{+0.20}_{-0.16}\times 10^{-5}$ eV$^2$ and $|\Delta m_{31}^2|=2.42^{+0.03}_{-0.04}\times 10^-3$ eV$^2$, for the inverted mass hierarchy. In the case of the normal ordering, we will use ... 

Let us divide our discussion in the different $U_{ij}(\phi,\sigma) U_{kl}(\phi',\sigma')$ combinations. The results of each case will be presented as follows: First, we give the full relations between the mixing angles, and $CP$ invariants, in terms of the correction parameters. Then, we present the plots which relate the mixing angles with the $CPV$ phases consistent with the experimental inputs. For the sake of simplicity, we will present only those plots where would be possible to restrict the allowed values of the $CPV$ phases in terms of one mixing angle\footnote{Plots where $CP$ phases cannot be bounded will not be presented but will be mentioned in the discussion.}. One possibility of restricting the $CP$ phases, even more, is to ask for small deviations from the $\mu-\tau$ symmetric mass matrix \cite{Rivera-Agudelo:2019seg,Rivera2}. This last requirement will be implemented by asking the breaking parameters in Eq. (\ref{deltaepsilon})  ($\hat \delta$ and $\hat \epsilon$) to be less than $0.3$. We will also include this additional restriction in each combination and will be shown when possible. The case where one of the correction phases is fixed to zero is also analyzed. Finally, to study the connection between the predicted phases and physical observables we also present the plot of neutrinoless double beta decay amplitude $(|m_{ee}|)$ versus the lightest neutrino mass ($m_0$) in both mass hierarchies.

\subsection{Case $U_{12}U_{13}$}
 As a first case let us consider a correction matrix of the form $U_{12}(\phi,\sigma) U_{13}(\phi',\sigma')$ in Eq. (\ref{Ucorr}). From Eq. (\ref{mixings}), we obtain the following expressions for the mixing angles in terms of the four correction parameters $(\phi^{(\prime)},\sigma^{(\prime)})$
\begin{eqnarray}\label{eq:mixing1213}
s^2\theta_{12} &=& -\frac{2 \left(\sqrt{2} c \sigma s 2\phi +2 s ^2\phi + c ^2\phi \right)}{s ^2\phi^\prime \left(-4 \sqrt{2} c \sigma s \phi c \phi - s ^2\phi + c ^2\phi + 3\right)-6}
\nonumber \\
s^2 \theta_{13} &=& \frac{1}{6} s ^2\phi^\prime \left(-2\sqrt{2} c\sigma s2\phi + c 2 \phi + 3\right)
\nonumber \\
s^2 \theta_{23} &=&\frac{1}{2 s^2\phi^\prime \left(-2 \sqrt{2} c\sigma s2\phi + c2\phi + 3 \right)-12} \nonumber  \\
&& \times \left[ -4 \sqrt{3} s\phi^\prime c\phi^\prime \left(\sqrt{2} s\phi c(\sigma^{\prime}-\sigma) + c\sigma^{\prime} c\phi \right) \right. \nonumber \\
&&~~~ \left. + s^2\phi^\prime \left(-2 \sqrt{2} c\sigma s2\phi + c2\phi - 3\right) - 6 c^2\phi^\prime  \right]~	.
\end{eqnarray}
It is important to note that relations in Eq. (\ref{eq:mixing1213}) change depending on the particular choice of correction matrices such that they need to be obtained in each case. Besides, the $CP$ invariants which give place to the $CPV$ phases are also expressed in terms of the correction parameters,
\begin{eqnarray}\label{eq:invariants1213}
J_{CP} &=& \frac{ s2\phi^\prime }{48 \sqrt{3}} \nonumber \\
&& \times \left[ c\phi \left(8 s^2\phi s(\sigma^{\prime} - 2\sigma) - 5s\sigma^{\prime} \right)+\sqrt{2}(s3\phi s(\sigma^{\prime}+\sigma) \right. \nonumber \\
&&~~~ \left. -3 s\sigma^{\prime} c\sigma s\phi + 5 c\sigma^{\prime} s\sigma s\phi )-3 s\sigma^{\prime} c3\phi \frac{}{} \right] 
\nonumber \\
I_1 &=& -\frac{2s \phi c^2\phi^\prime }{9} \left(s4\sigma s^3\phi +\sqrt{2} s\sigma c^3\phi - 3 s\sigma c\sigma s\phi c^2\phi - \sqrt{2} s3\sigma s^2\phi c\phi \right)
\nonumber \\
I_2 &=&\frac{ s2\sigma^\prime s^22\phi^\prime}{288} \nonumber \\
&& \times \left(-8 c2\sigma s^22\phi + 4 \sqrt{2} c\sigma (6 s2\phi + s4\phi ) - 12 c2\phi + 3 c4\phi - 23\right) ~.
\end{eqnarray}

\begin{figure}\centering
\includegraphics[scale=0.4]{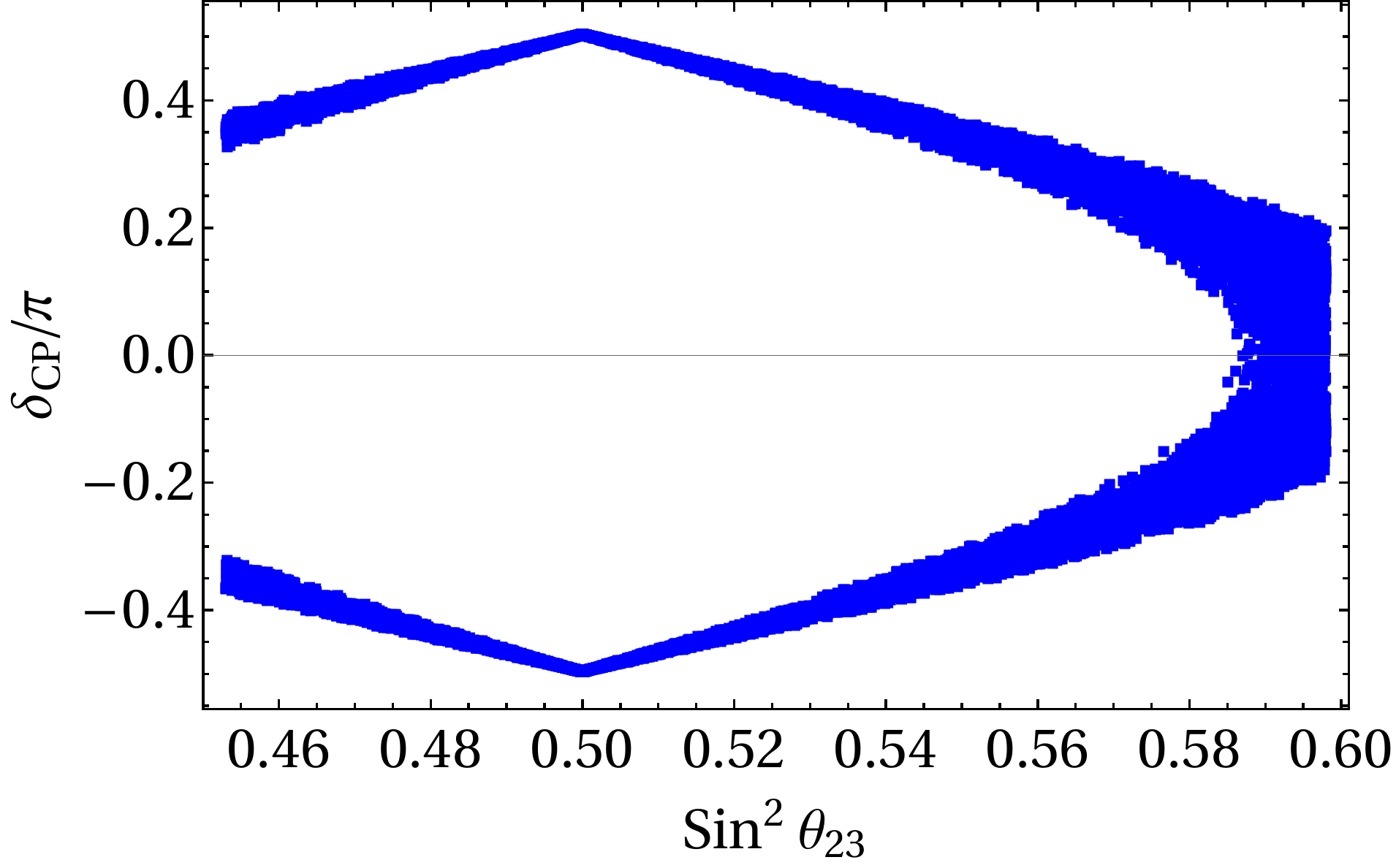} 
\includegraphics[scale=0.4]{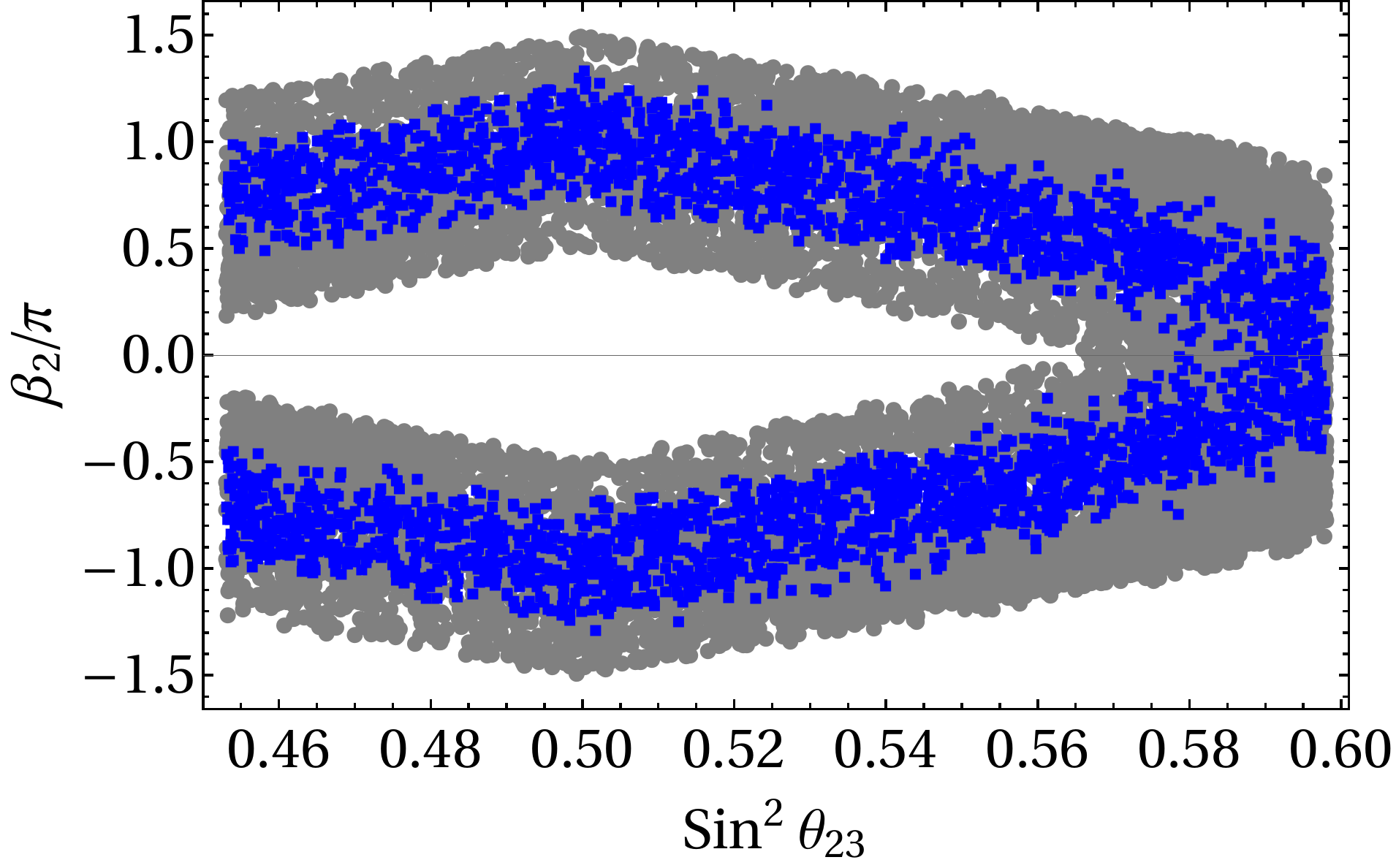}
\caption{Allowed region of $CP$ phases $\delta_{CP}$ and $\beta_2$ in terms of the atmospheric angle for the $U_{12}(\phi,\sigma) U_{13}(\phi',\sigma')$ combination. Gray points (blue squares) region is obtained when the restriction of small deviations in the mass matrix is omitted (included).}\label{fig:12-13-libres}
\end{figure}

To proceed with our analysis, we vary the correction parameters in the full $(-\pi,\pi)$ range and pick up those values consistent with the $3\sigma$ ranges of mixing angles. Then, we can substitute the allowed values in the expressions of $CP$ invariants of Eq. (\ref{eq:invariants1213}) and look for regions relating $CPV$ phases of the mixing matrix with at least one mixing angle. In Fig. \ref{fig:12-13-libres} we present the plot of the Dirac ($\delta$) and Majorana ($\beta_2$) phases related with $\sin^2 \theta_{23}$. It is interesting to observe that such phases present noticeable correlations with this mixing angle and could be restricted with an improved determination of $\theta_{23}$. We do not show the region of $\beta_1$ since it cannot be restricted by any of the mixings and remains arbitrary in any case. We can also see that if we ask for small deviations from $\mu-\tau$ symmetry in the mass matrix, the region of the Majorana phase is additionally restricted (blue region), while $\delta_{CP}$ remains unchanged. These regions show, for instance, that $\delta_{CP}$ and $\beta_2$ are different from zero when $\theta_{23}$ has the TBM value, but a value near its upper $3\sigma$ limit is in favor of null (or small) values of these phases.  

\begin{figure}\centering
\includegraphics[scale=0.4]{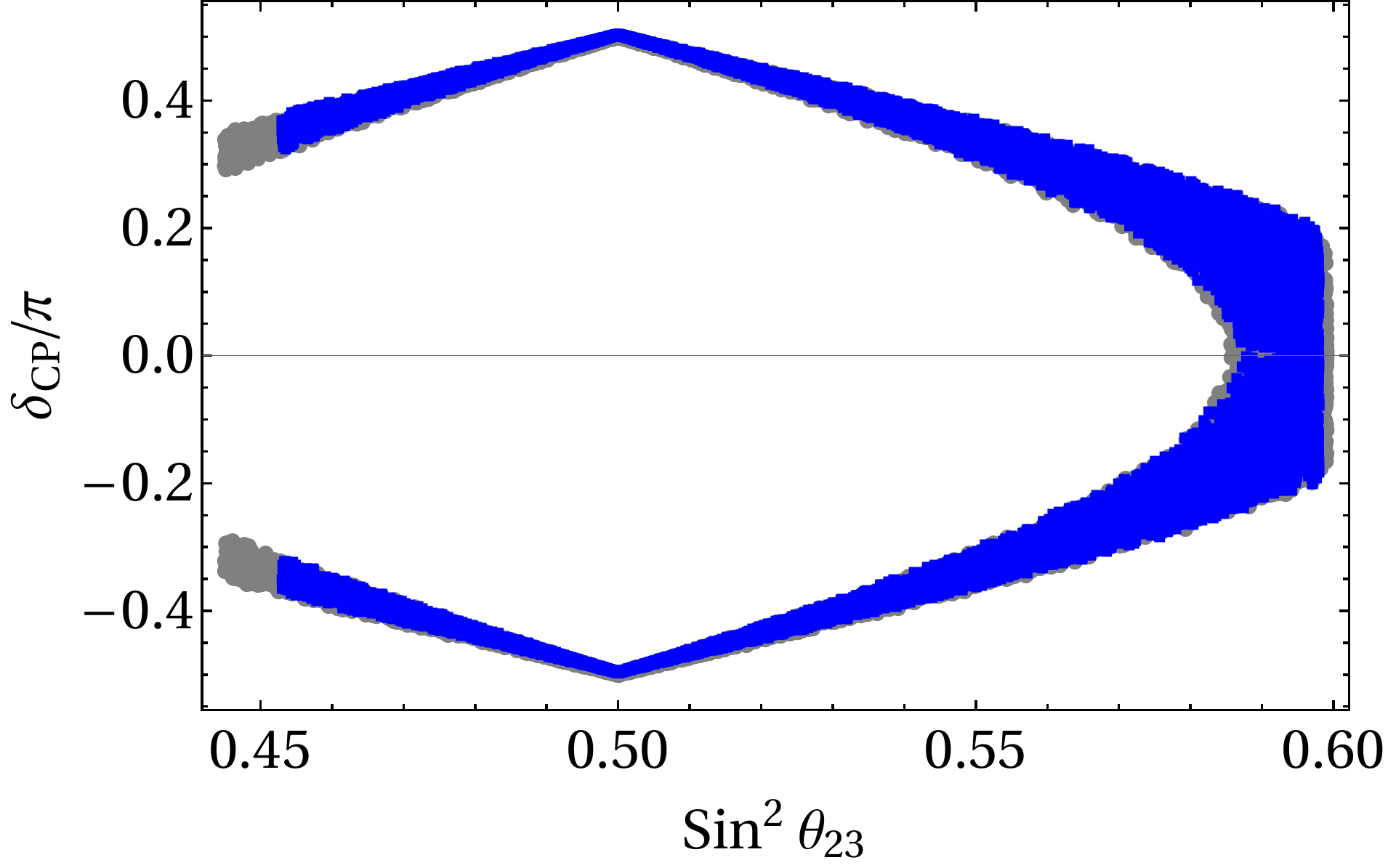} 
\includegraphics[scale=0.4]{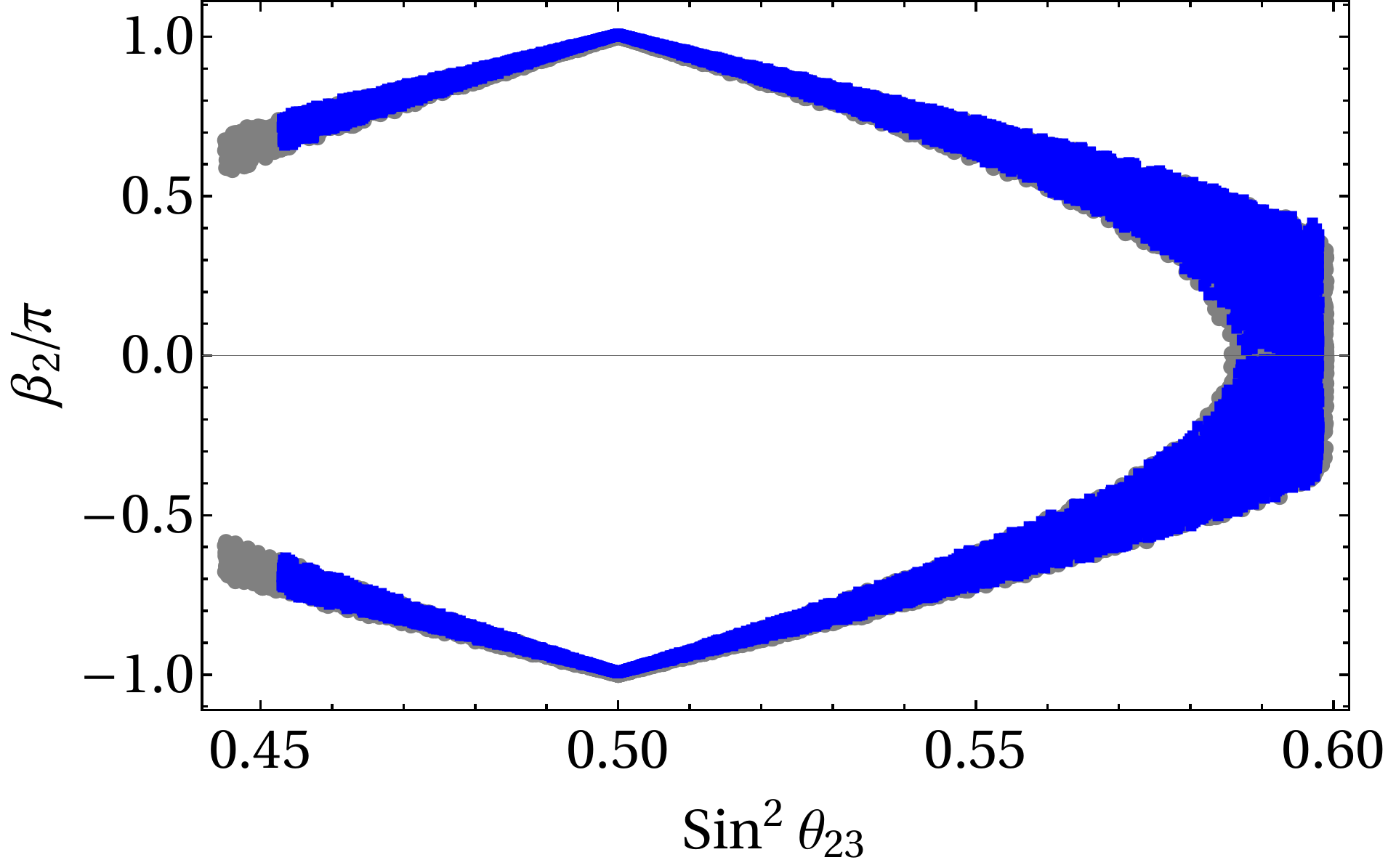}
\includegraphics[scale=0.4]{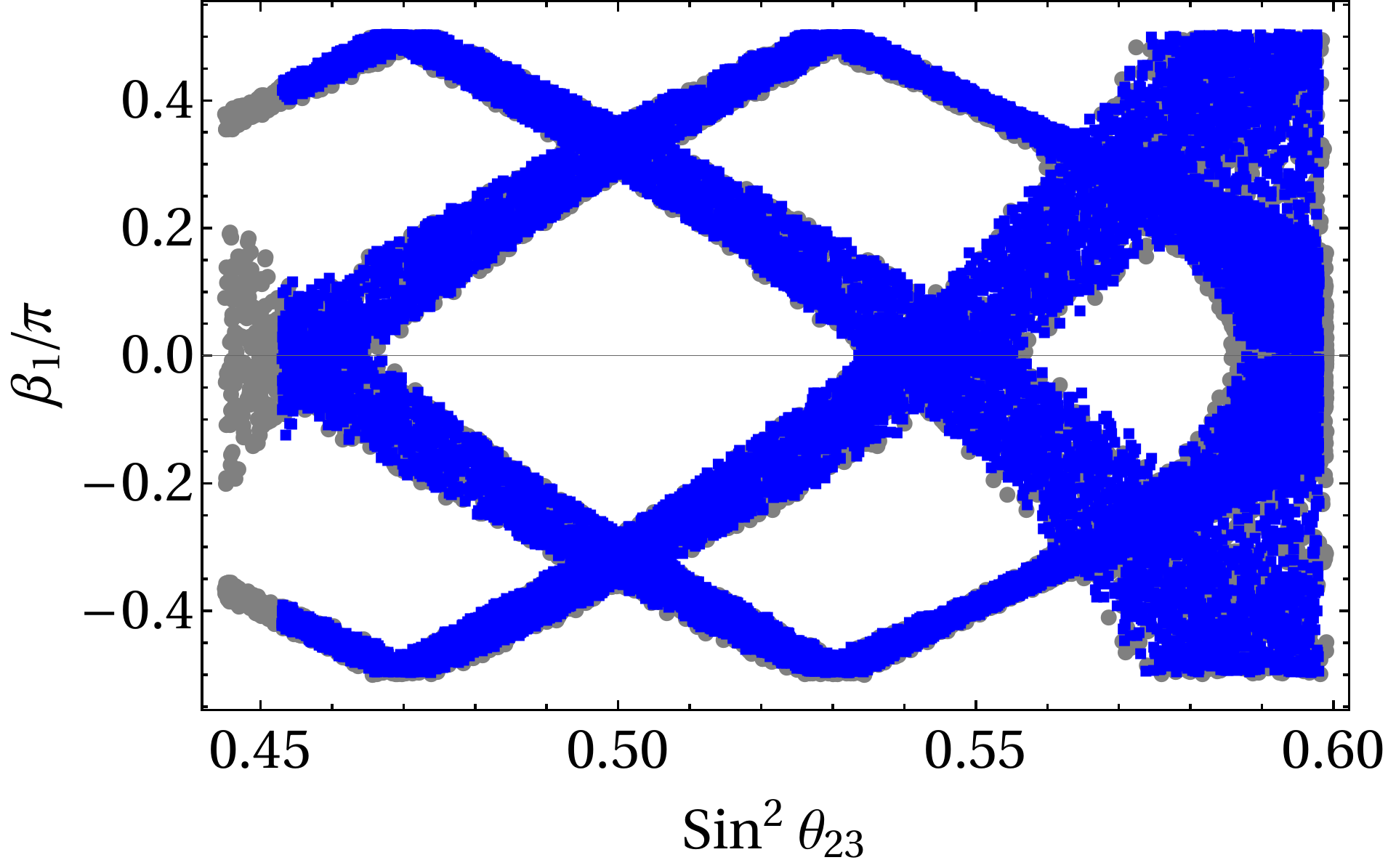}
\caption{Same description as in Fig. \ref{fig:12-13-libres} for the $\sigma' = 0$ case.}\label{fig:12-13-cerolibre}
\end{figure}

\begin{figure}\centering
\includegraphics[scale=0.4]{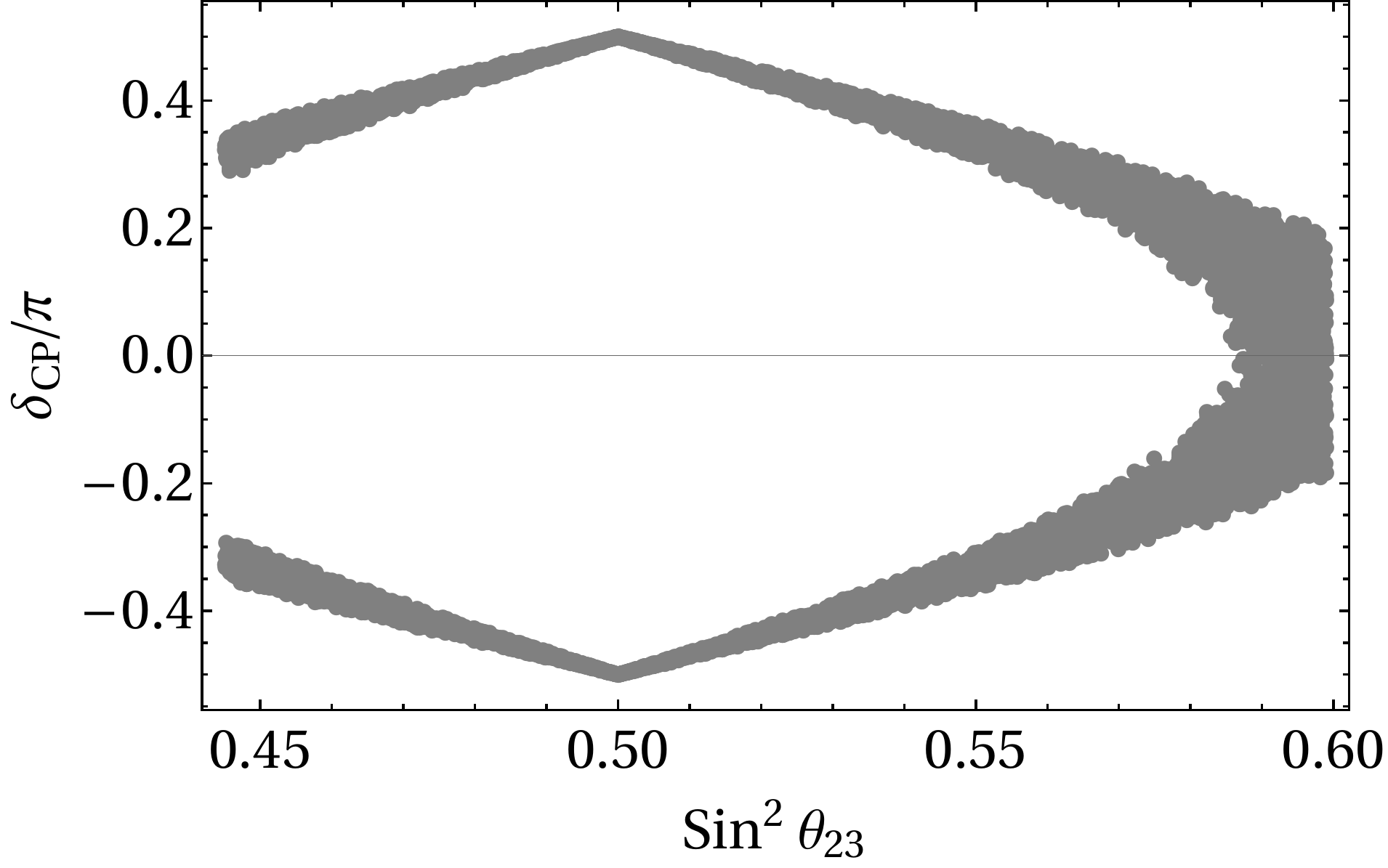} 
\includegraphics[scale=0.4]{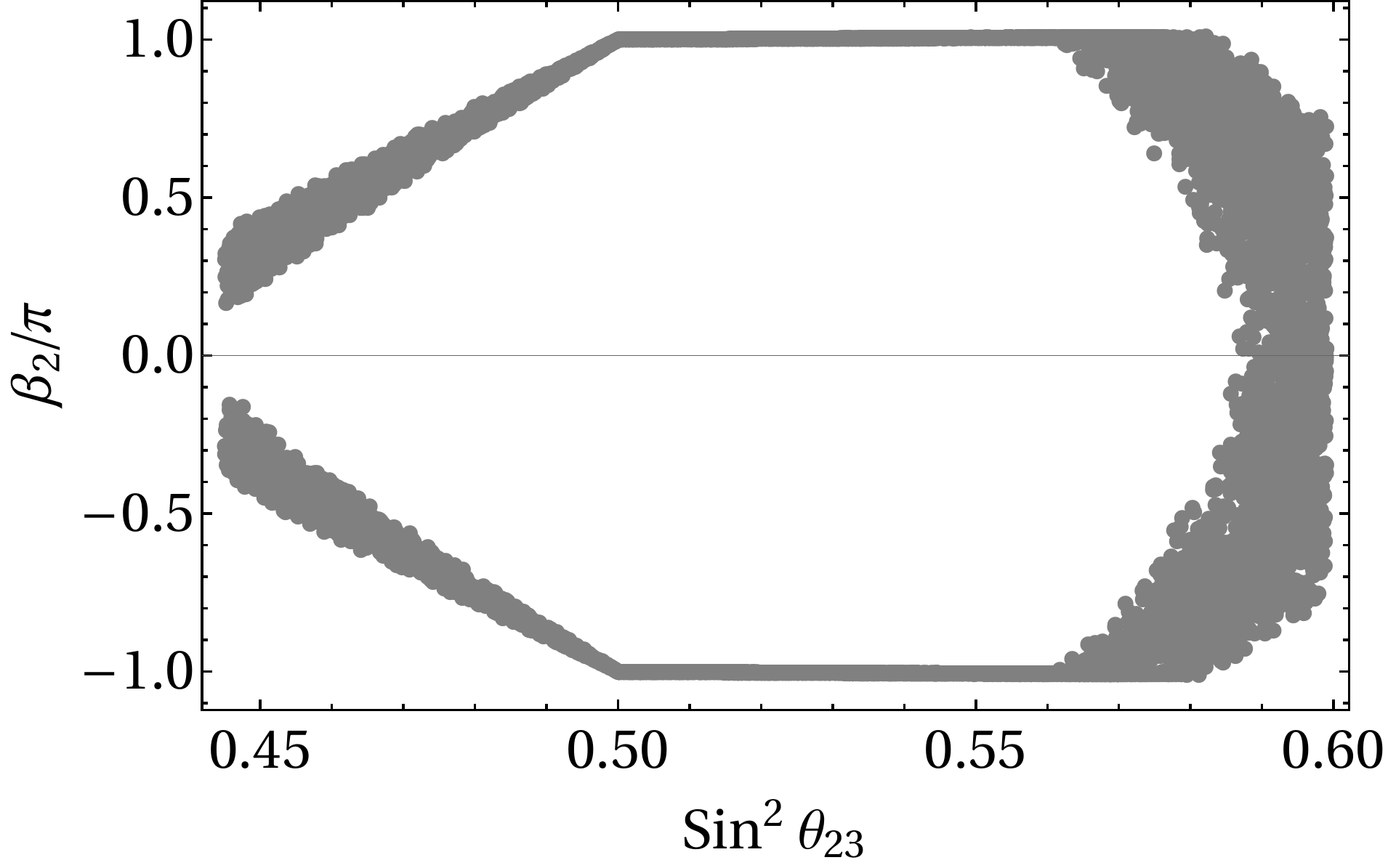}
\caption{Same description as in Fig. \ref{fig:12-13-libres} for the $\sigma=0$ case.}\label{fig:12-13-librecero}
\end{figure}

In Fig. \ref{fig:12-13-cerolibre} we show the relation between the three $CPV$ phases and the mixing angle in the case $\sigma' = 0$. We observe in this case that it is possible to bound, indirectly, all the three $CP$ phases through the atmospheric angle as marked relations are obtained. From the third term in Eq. \ref{eq:invariants1213}, it is straightforward to show that $\delta$ and $\beta_2$ are directly linked each other because of $\sin(\beta_2 + 2\delta_{CP}) = 0$, which can be seen in Fig. \ref{fig:12-13-cerolibre}. The region is also slightly reduced when the condition of small deviations in the mass matrix is impossed. 

For the selection $\sigma=0$ we obtain striking regions for the $\delta$ and $\beta_2$ phases which are shown in Fig. \ref{fig:12-13-librecero}. Contrary to the previous case, $\beta_1$ is fixed to zero as can be readily verify from second expression in Eq. (\ref{eq:invariants1213}) by setting $\sigma = 0$. Finally, the present scheme forbids the inclusion of small deviations in the mass matrix such that no regions are obtained, which is consistent with previous studies where $\beta_1$ is set to zero \cite{Rivera2}.    

\begin{figure}\centering
\includegraphics[scale=0.6]{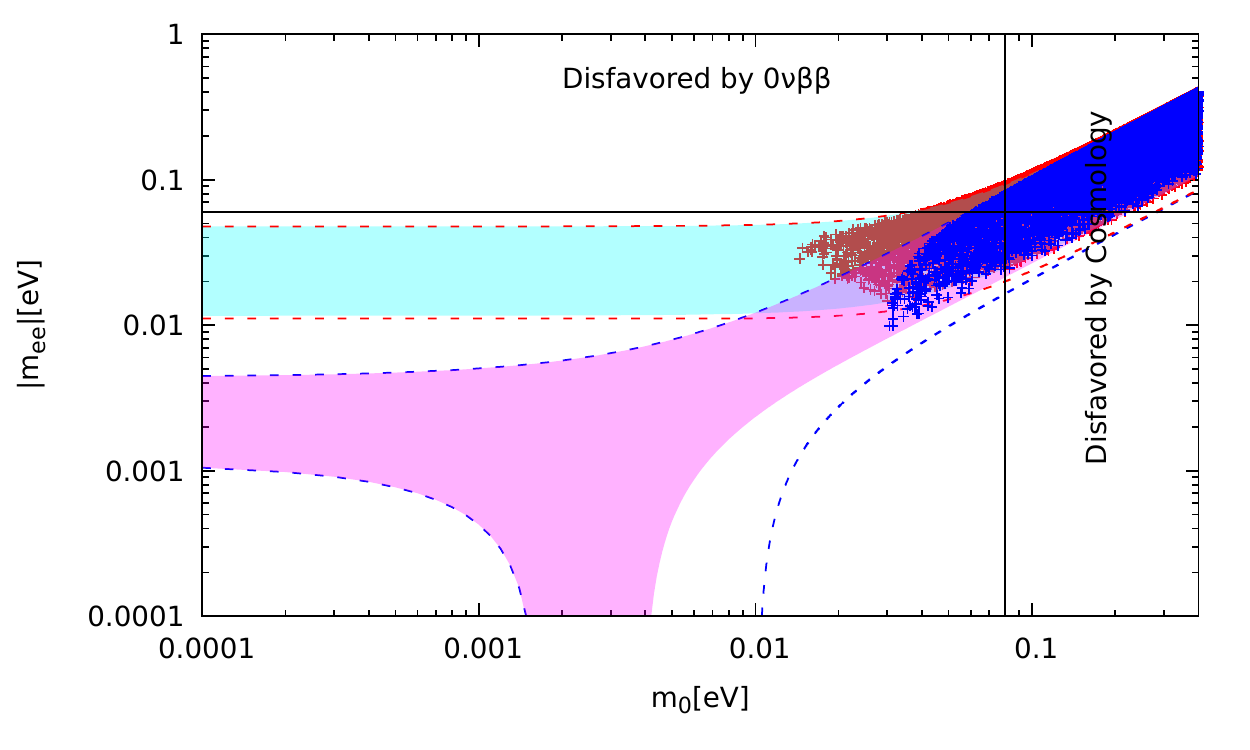} 
\includegraphics[scale=0.6]{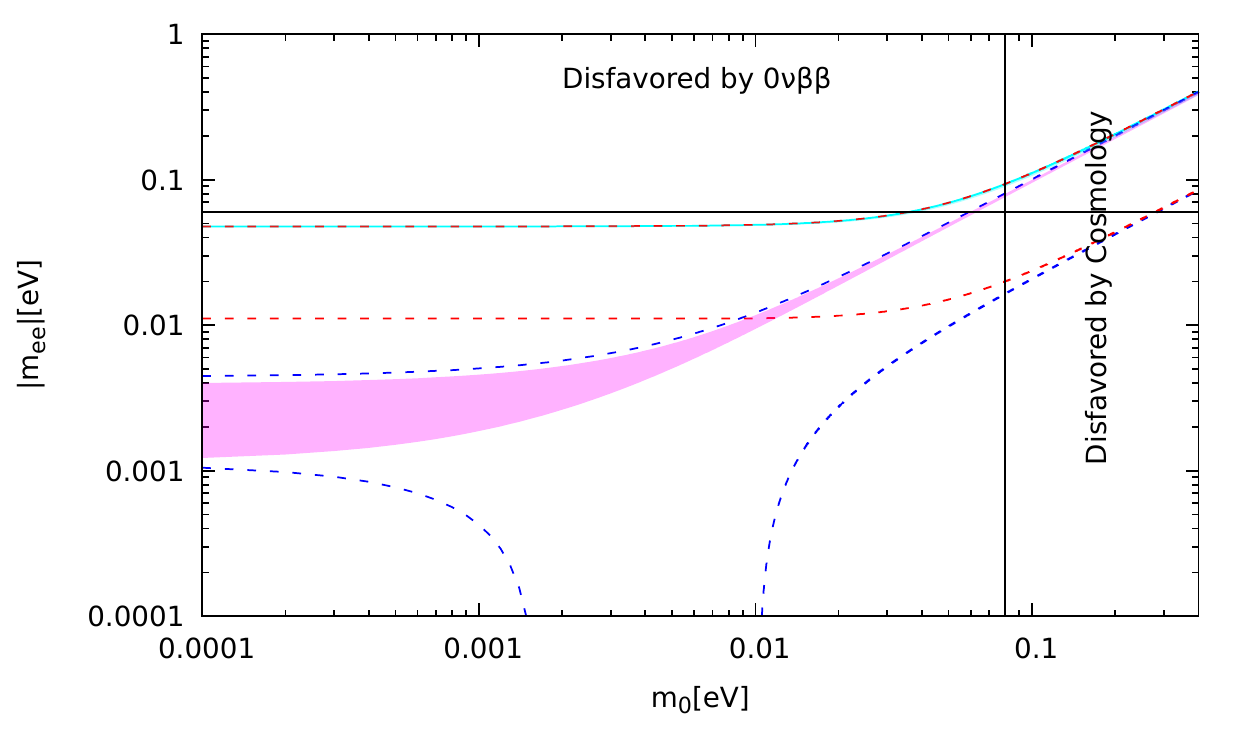}
\caption{Predicted regions of $|m_{ee}|$ for the $U_{12}(\phi,\sigma) U_{13}(\phi',\sigma')$ and $\sigma'=0$  cases (left) and $\sigma=0$ (right). Region delimited by dotted red (blue) lines corresponds to the IH (NH). Magenta (cyan) solid region shows the allowed region by the predicted $CPV$ phases in the NH (IH). Scattered blue (red) crosses denote the allowed region when the consideration of small breaking in the mass matrix is included in the NH (IH). Experimental limits on $|m_{ee}|$ and $m_0$ are taken from \cite{Auger:2012ar,Albert:2014awa,Gando:2012zm,Ade:2015xua}. }\label{fig:12-13-0bb}
\end{figure}

Neutrinoless double beta decay amplitude $|m_{ee}|$ may serve as a possible way of testing the above scenarios. As can be observed in Fig. \ref{fig:12-13-0bb}, the predicted region for this amplitude shrinks for a correction of the form $U_{12}(\phi,\sigma) U_{13}(\phi',\sigma')$ in Eq. (\ref{eq:corr}) in comparison to the one obtained from unbounded $CP$ phases (Eq. \ref{PMNS}). Such a reduction can be understood from the emergent correlations between the $CP$ invariants given by the correction parameters. We can observe that when small deviations from the $\mu-\tau$ symmetric mass matrix are required in the general and $\sigma^\prime=0$ cases, these regions are reduced to the quasi-degenerate mass hierarchy regime, which is consistent with previous parametrization independent analysis \cite{Gupta:2013it, Rivera2}. For these last combinations, the predicted regions of $|m_{ee}|$ are indistinguishable between them since terms in the amplitude involving $\sigma'$ are suppressed by contributions of the form $\sin^2 \phi \sim 0.004$ \footnote{However, for the sake of space, complete expressions of $|m_{ee}|$ in terms of the four correction parameters will not be shown in the remainder sections, but could be obtained, after some algebra, from the $ee$ entry of $M_{\nu}=U_{PMNS}{\rm diag}(m_1,m_2,m_3) U_{PMNS}^{\rm T}$, with $U_{PMNS}$ given in Eq. (\ref{eq:corr}).}, hence, only one plot is presented for both cases. For the $\sigma=0$ selection, the allowed regions are narrow as a consequence of $\beta_1=0$. In the IH we can observe that the predicted region is near the reach of $0\nu\beta\beta$ experiments such that could be easily tested. Therefore, forthcoming bounds in $|m_{ee}|$, in addition to more precise determinations of mixing angles, could help to discard some of the considered schemes.

%%%%%%%%%%%%%%%%%%%%%%%%%%%%%%%%%%%%%%%%%%%%%%%%%%%%%%%%%%%%%%%%%%%%%%%%%%%%%%%%%%%%%%%%%%%%%%%%%%%%%%%%%%%%%%%%%%%%%%%%%%%%%%%
%%%%%%%%%%%%%%%%%%%%%%%%%%%%%%%%%%%%%%%%%%%%%%%%%%%%%%%%%%%%%%%%%%%%%%%%%%%%%%%%%

\subsection{Case $U_{12}U_{23}$}

 When the correction matrix in Eq. (\ref{Ucorr}) takes the form $U_{12}(\phi,\sigma) U_{23}(\phi',\sigma')$, the relations between mixing angles and correction parameters read
\begin{eqnarray}\label{eq:mixings-1223}
s^2 \theta_{12} &=& \frac{c^2\phi^\prime \left(\sqrt{2} c\sigma s2\phi + 2s^2\phi + c^2\phi\right)}{3 \left(1-\frac{1}{3} s^2\phi^\prime \left(s^2\sigma c^2\phi + \left(c\sigma c\phi +\sqrt{2} s\phi \right)^2\right)\right)}
\nonumber \\
s^2 \theta_{13} &=& -\frac{1}{6} s^2\phi^\prime \left(-2 \sqrt{2} c\sigma s2\phi + c2 \phi - 3\right)
\nonumber \\
s^2 \theta_{23} &=& \frac{1}{2 s^2\phi^\prime \left(-2 \sqrt{2} c\sigma s2\phi + c2\phi - 3\right) + 12}  \nonumber \\
&& \times \left [ 4 \sqrt{3} s\phi^\prime c\phi^\prime \left(s\phi c(\sigma+\sigma^\prime)-\sqrt{2} c\sigma^\prime c\phi\right) \right. \nonumber \\
&&~~~ \left. + s^2\phi^\prime \left(-2 \sqrt{2} c\sigma s2 \phi + c2\phi + 3\right)+6 c^2\phi^\prime \right] ~,
\end{eqnarray}
while the $CP$ invariats are of the form
\begin{eqnarray}\label{eq:invariants-1223}
J_{CP} &=& -\frac{s2\phi^\prime}{48 \sqrt{3}} \nonumber \\
&& \times  \left[ s\sigma c\sigma^\prime (s3\phi - 7s\phi)-\sqrt{2} c\phi (s(2\sigma+\sigma^\prime)-4 s\sigma^\prime) \right.  \nonumber \\ 
&& ~~~\left. + c\sigma s\sigma^\prime (5 s3\phi - 3s\phi)+\sqrt{2} c3\phi s(2 \sigma+\sigma^\prime)\right]
\nonumber \\
I_1 &=& -\frac{2}{9} c^2\phi^\prime s\phi \nonumber \\
&& \times \left(s4\sigma s^3\phi + \sqrt{2} s\sigma c^3\phi - 3 s\sigma c\sigma s\phi c^2\phi -\sqrt{2} s3\sigma s^2\phi c\phi \right)
\nonumber \\
I_2 &=& \frac{1}{72} s^2\phi^\prime \left[ 8 c2(\sigma+\sigma^\prime) \left(\sqrt{2} s\sigma s2\phi + 2 s2\sigma c2\phi \right)  \right. \nonumber \\
&& \left. - s2(\sigma+\sigma^\prime) \left(4 c2\sigma (c4\phi + 3) + 4 \sqrt{2} c\sigma s4\phi + 3 c4\phi - 3\right) \right] ~.
\end{eqnarray}

\begin{figure}\centering
\includegraphics[scale=0.4]{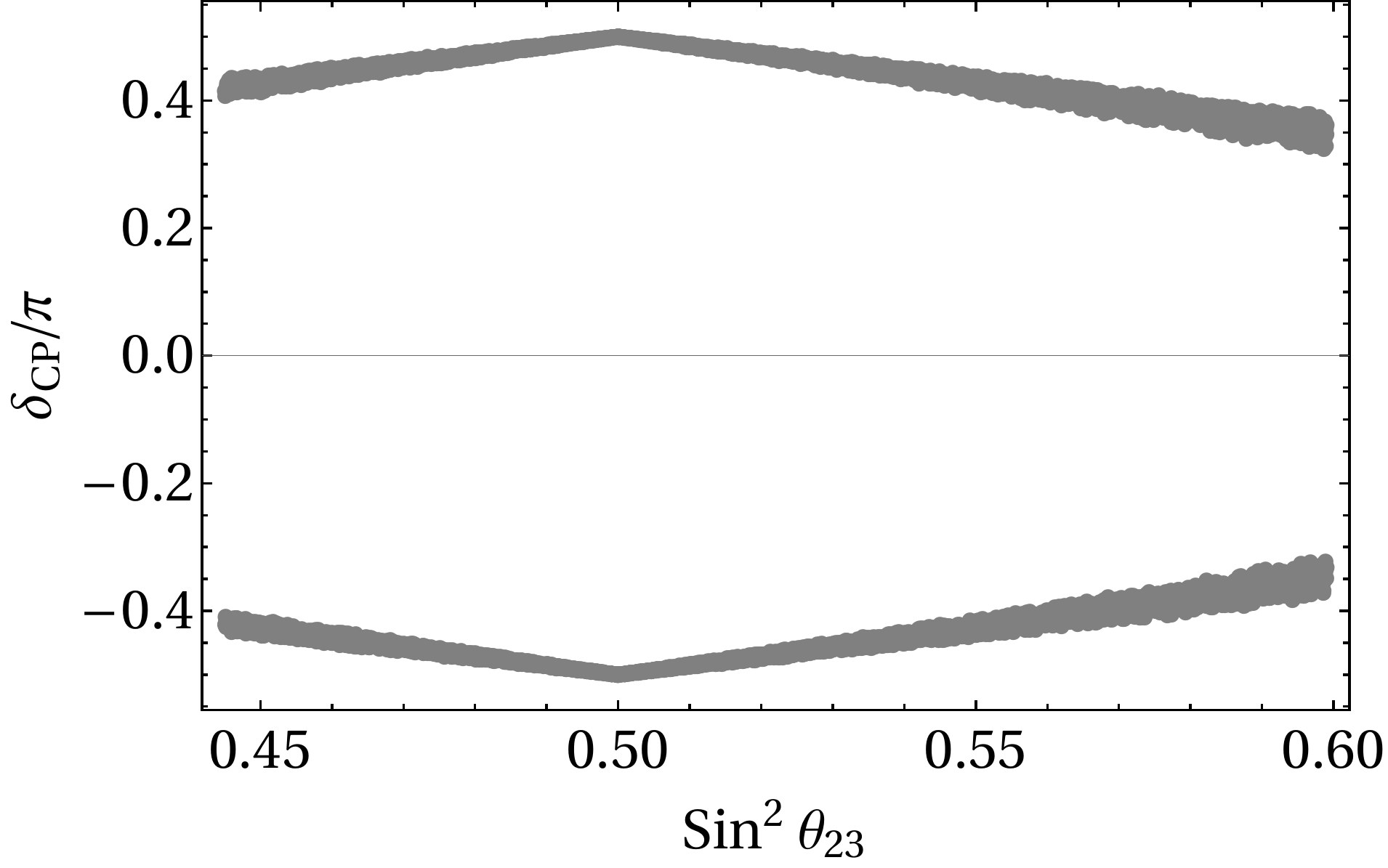} 
\includegraphics[scale=0.4]{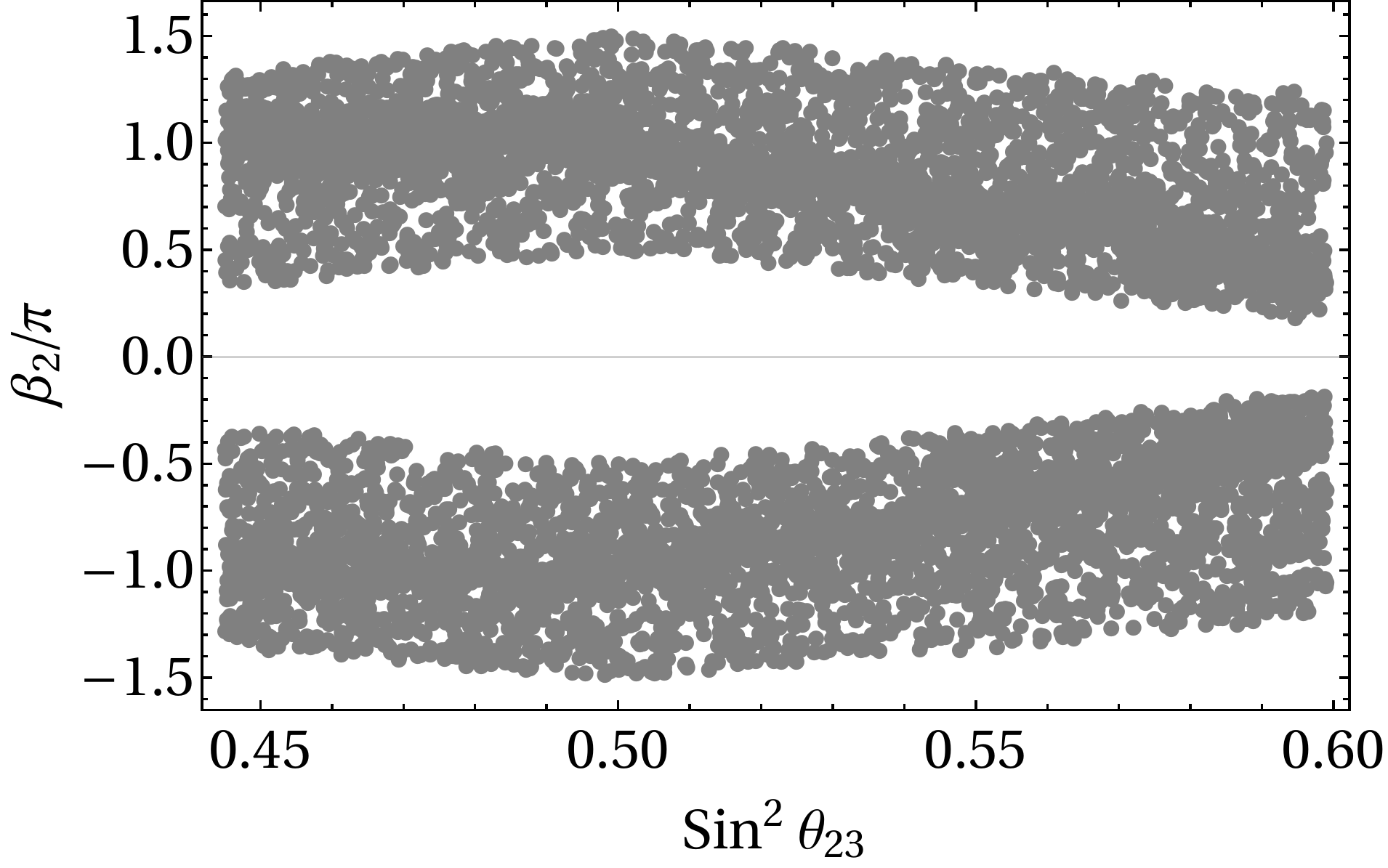}
\caption{Same description as in Fig. \ref{fig:12-13-libres} for the $U_{12}(\phi,\sigma) U_{23}(\phi',\sigma')$ case.}\label{fig:12-23-libres}
\end{figure}

 In Fig. \ref{fig:12-23-libres}, we plot the $\delta_{CP}$ and $\beta_2$ phases related to the $\sin^2 \theta_{23}$. Sharpness of $\delta_{CP}$ is understood from the smallness of $\theta_{13}$ since both are linked \textit{via} the parameter $\phi'$, as can be checked from Eqs. (\ref{eq:mixings-1223}) and (\ref{eq:invariants-1223}). There is no visible corelation between $\beta_1$ and the mixing angles since no major restrictions are imposed over $\phi$. Also, this combination is not compatible with the requierement of small departures from $\mu-\tau$ symmetry in the mass matrix. 

\begin{figure}\centering
\includegraphics[scale=0.4]{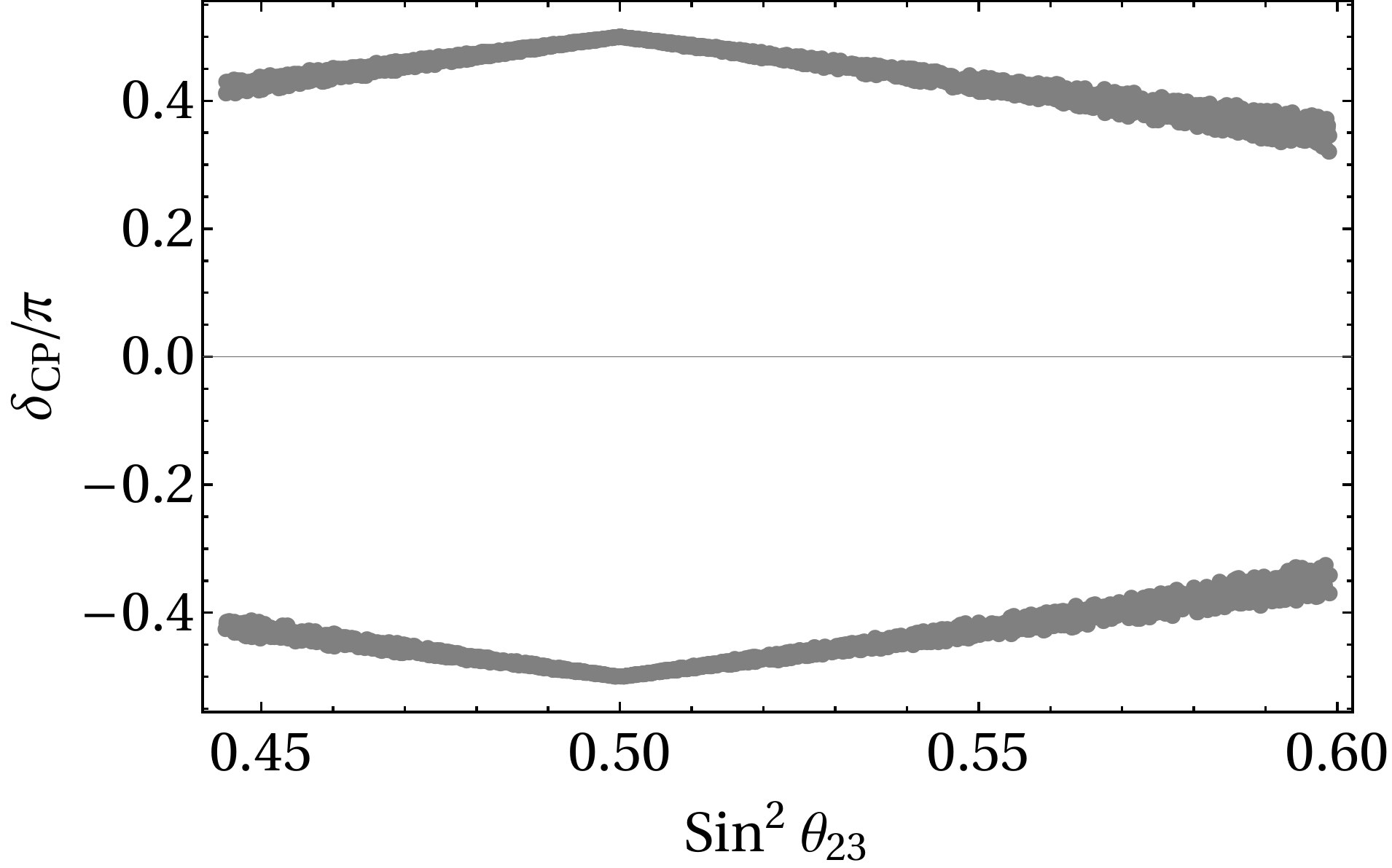} 
\includegraphics[scale=0.4]{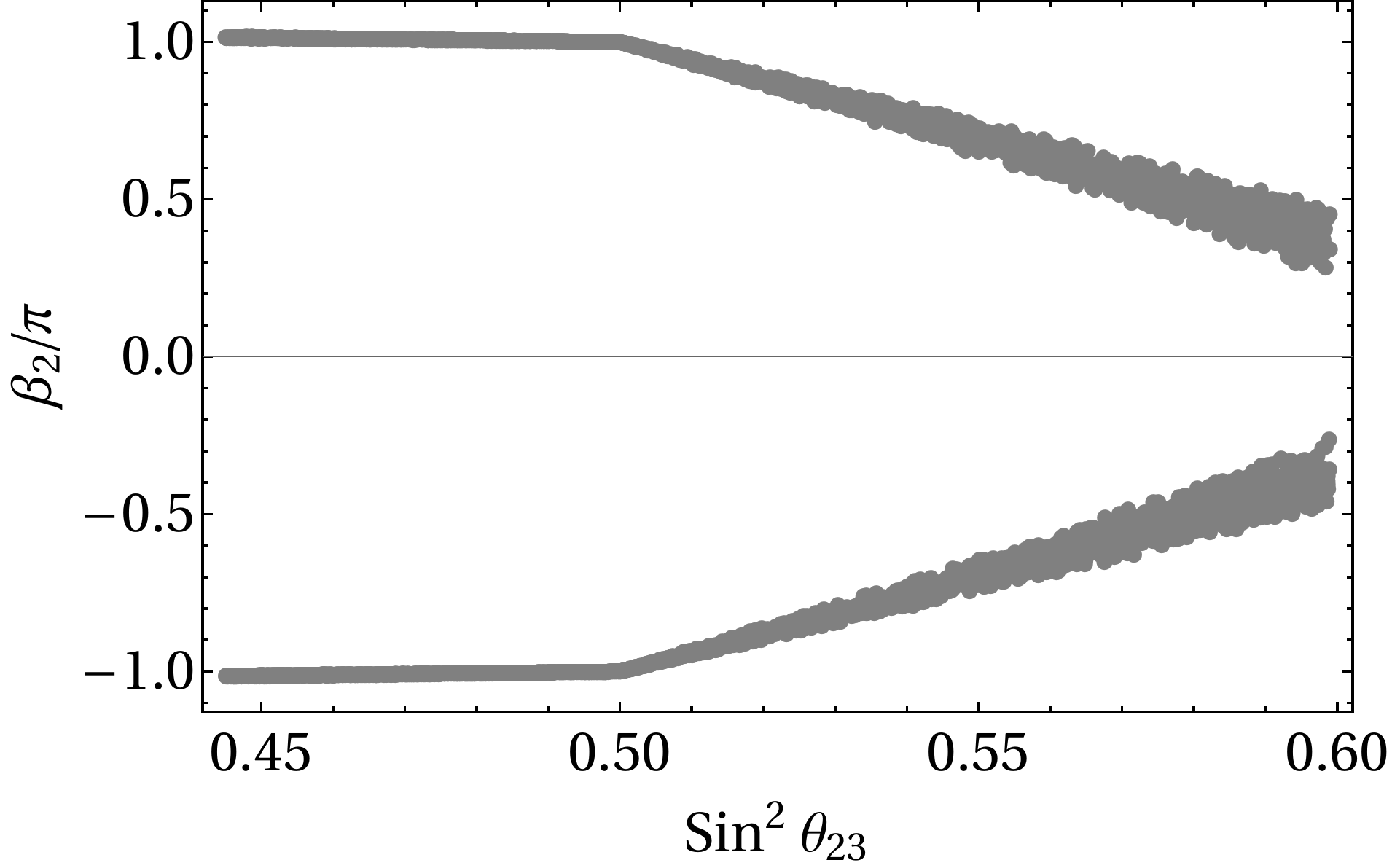}
\caption{Same description as in Fig. \ref{fig:12-13-libres} for the case $\sigma$ = 0.}\label{fig:12-23-cerolibre}
\end{figure}

For $\sigma=0$, $\beta_1$ is fixed to zero, as can be seen from $I_1$ in Eq. (\ref{eq:invariants-1223}). Regions for $\delta_{CP}$ and $\beta_2$ are shown in Fig. \ref{fig:12-23-cerolibre}, they are different from zero in the full range of $\sin^2 \theta_{23}$ and are strongly linked to this mixing angle. 

For the selection $\sigma'=0$, no regions compatible with experimental neutrino mixings were obtained such that this scenario is totally ruled out.  

\begin{figure}\centering
\includegraphics[scale=0.6]{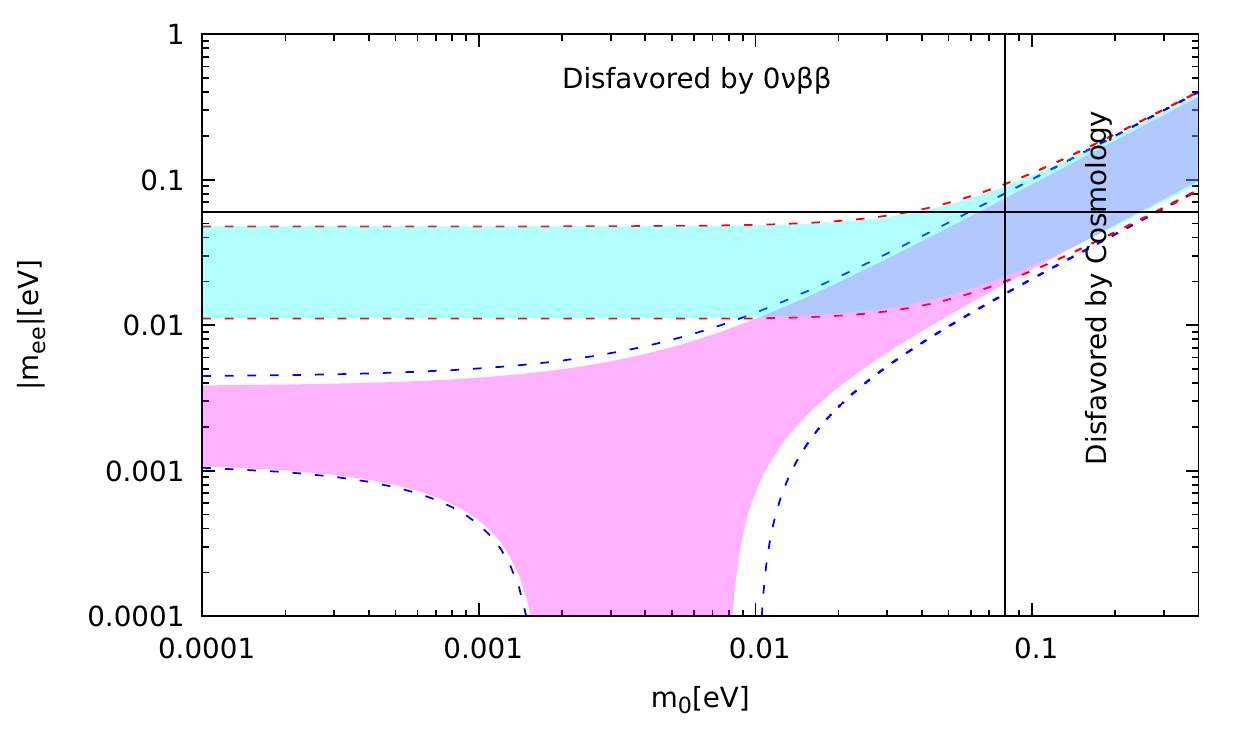} 
\includegraphics[scale=0.6]{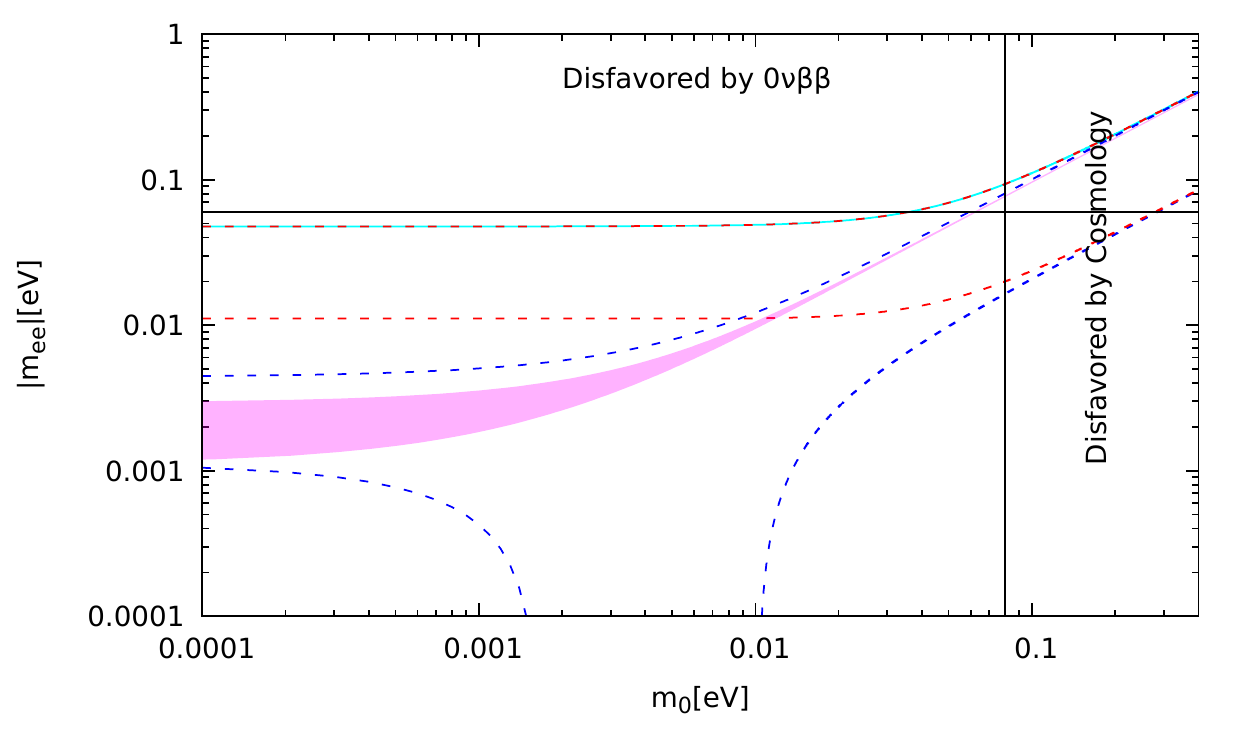}
\caption{Same description as in Fig. \ref{fig:12-13-0bb} but for $U_{12}(\phi,\sigma) U_{23}(\phi',\sigma')$ ~(left) and $\sigma = 0$ ~(right).}\label{fig:12-23-0bb}
\end{figure}

Regions of the neutrinoless double beta decay amplitude are shown in Fig. \ref{fig:12-23-0bb} for the general case $U_{12}(\phi,\sigma) U_{23}(\phi',\sigma')$ and $\sigma=0$. Predicted regions are marked for $\sigma=0$ given the sharp values of the $CP$ phases obtained in this case.

%%%%%%%%%%%%%%%%%%%%%%%%%%%%%%%%%%%%%%%%%%%%%%%%%%%%%%%%%%%%%%%%%%%%%%%%%%%%%%%%%%%%%%%%%%%%%%%%%%%%%%%%%%%%%%%%%%%%%%%%%%%%%%%
%%%%%%%%%%%%%%%%%%%%%%%%%%%%%%%%%%%%%%%%%%%%%%%%%%%%%%%%%%%%%%%%%%%%%%%%%%%%%%%%%%%%%%%%%%%%%%%%%%%%%%%%%%%%%%%%%%%%%%%%%%%%%

\subsection{Case $U_{13}U_{12}$}

\begin{figure}\centering
\includegraphics[scale=0.4]{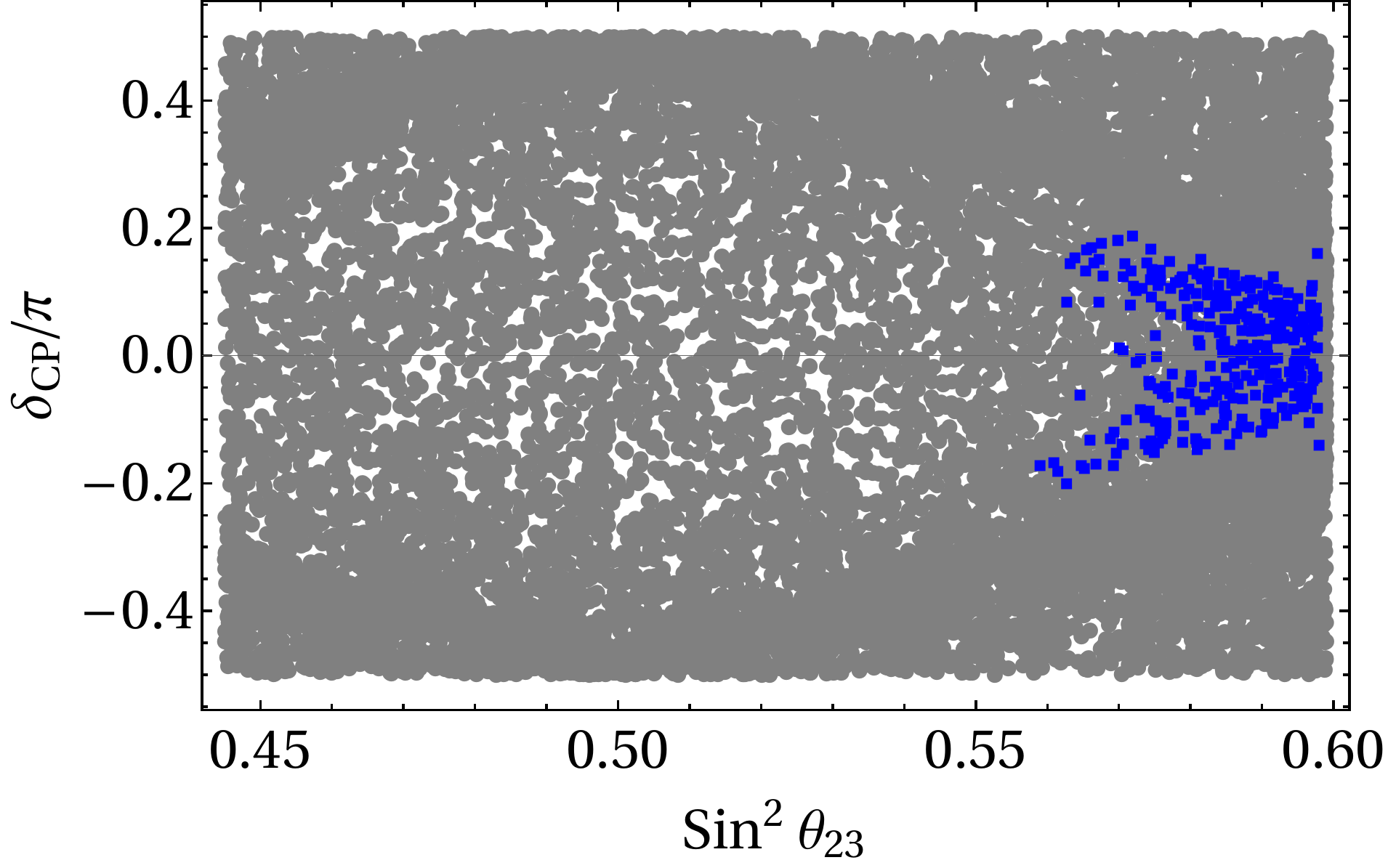}
\includegraphics[scale=0.4]{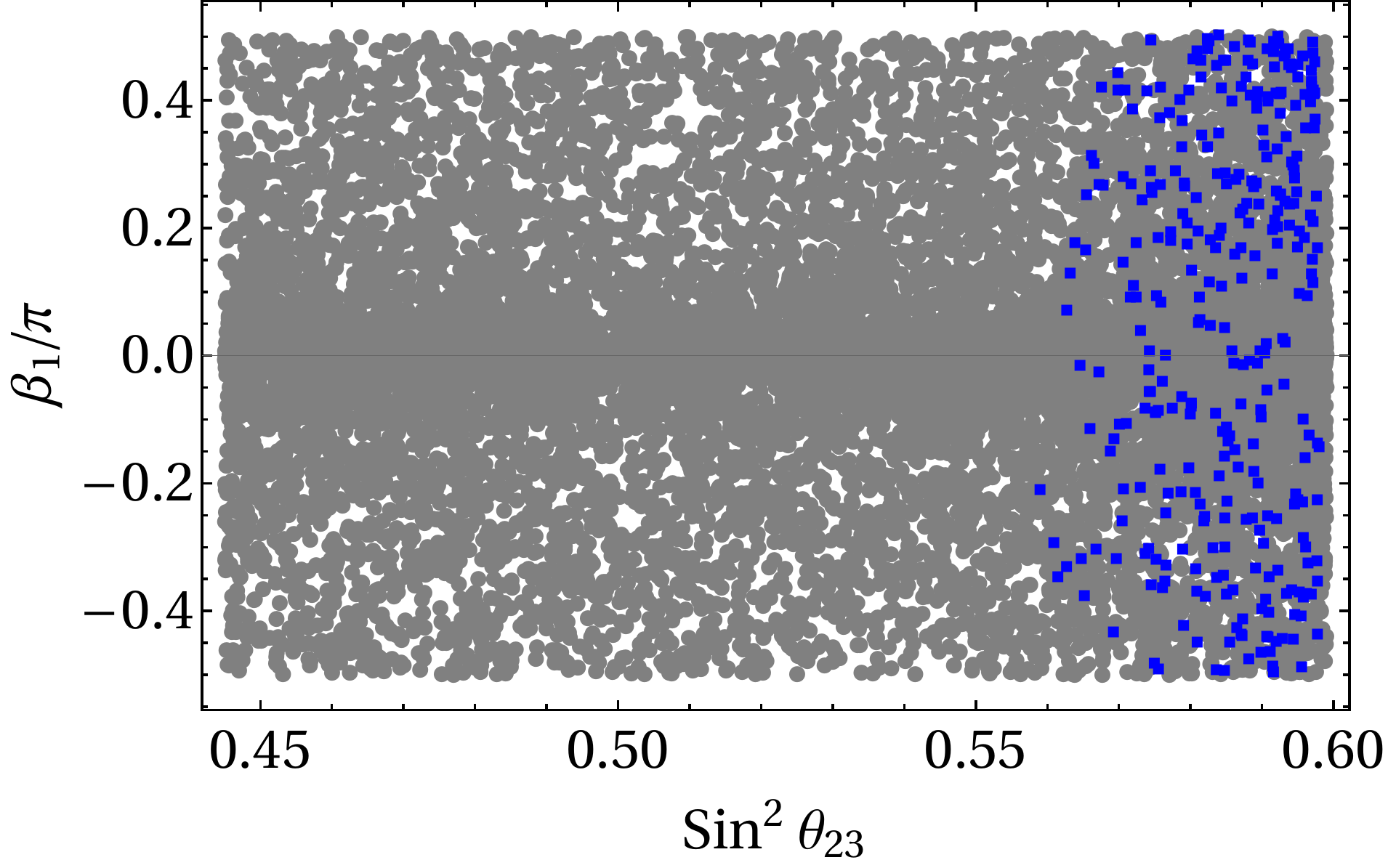} 
\includegraphics[scale=0.4]{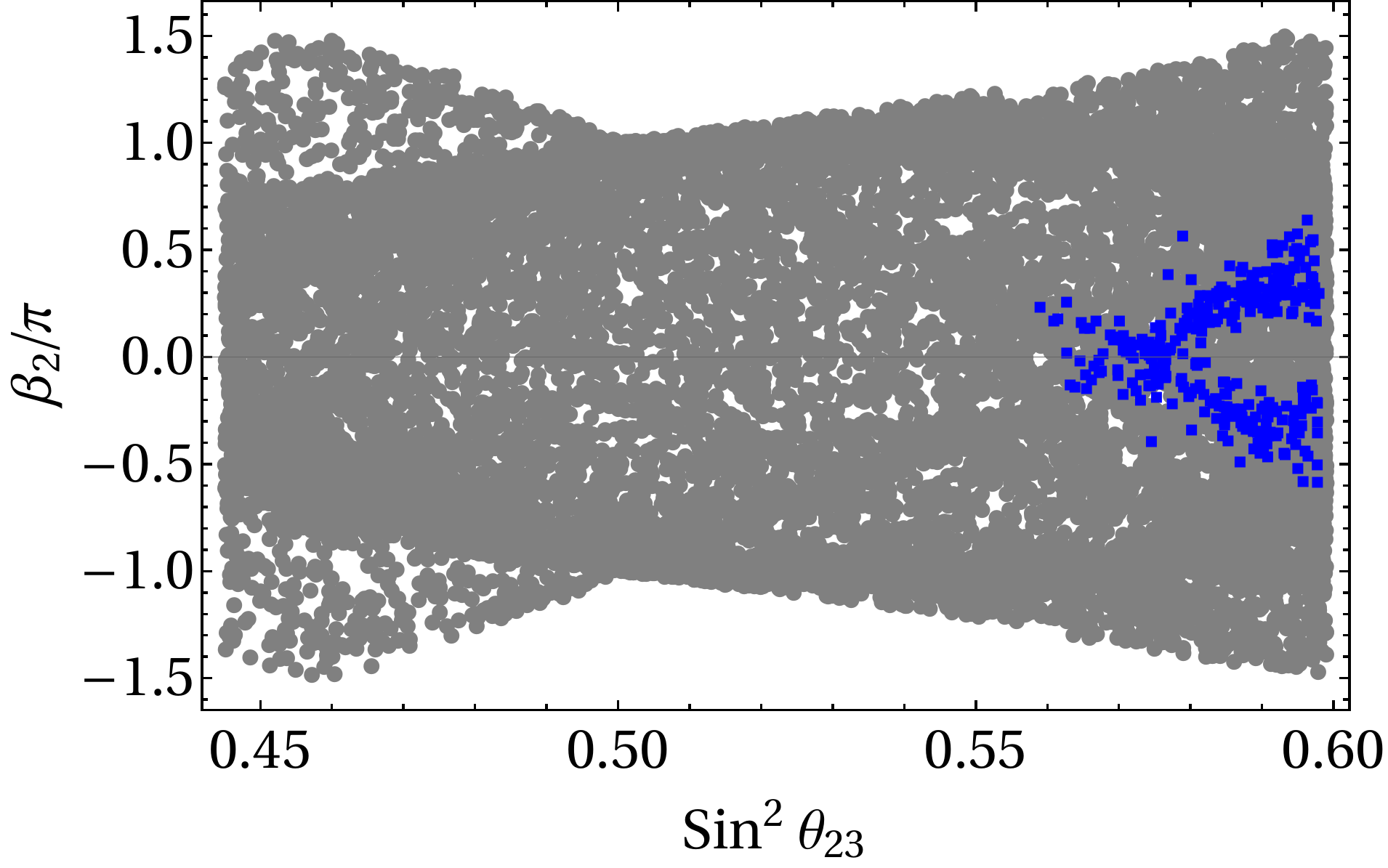}
\caption{Same description as in Fig. \ref{fig:12-13-libres} for the $U_{13}(\phi,\sigma) U_{12}(\phi',\sigma')$ case.}\label{fig:13-12-libres}
\end{figure}

The expressions of the mixing angles in terms of the correction parameters for the combination $U_{13}(\phi,\sigma)U_{12}(\phi',\sigma')$ are given by
\begin{eqnarray}\label{eq:mixings1312}
s^2 \theta_{12} &=& \frac{\sqrt{2} c\sigma^\prime s2\phi^\prime c\phi + s^2\phi^\prime c2\phi + 1}{c2\phi + 2} \nonumber \\
s^2 \theta_{13}  &=& \frac{2 s^2\phi}{3} \nonumber \\
s^2 \theta_{23} &=& \frac{1}{2} \left(\frac{\sqrt{3} c\sigma s2\phi}{c2\phi + 2}+1\right)~ .
\end{eqnarray}
From the second term in Eq. (\ref{eq:mixings1312}) we should expect that the $\phi$ parameter be restricted to small values because of the reactor angle ($\theta_{13}$), which also leads to small deviations of the atmospheric angle from its maximal value $\theta_{23}\sim \pi/4$. On the contrary, the correction parameter $\phi'$ is less restricted since it is only related to the solar angle. For the $CPV$ phases, one has the expressions
\begin{eqnarray}
J_{CP} &=& \frac{\sqrt{2} s2\phi^\prime \left( s\sigma c\sigma^\prime (s\phi - s3\phi)+2 c\sigma s\sigma^\prime s3\phi \right)-4 s\sigma c2\phi^\prime s2\phi}{24 \sqrt{3}} \nonumber \\
I_1 &=& -\frac{s\sigma^\prime s\phi^\prime}{9} \left[ \frac{}{} 8 c\sigma^\prime c2\sigma^\prime s^3\phi^\prime c^2\phi + c\sigma^\prime s\phi^\prime c^2\phi^\prime (-4 c2\phi + c4\phi-3) \right. \nonumber \\
&&\left.  ~~~ - \sqrt{2}  s^2\phi^\prime c\phi^\prime  (2 c2\sigma^\prime + 1)(c\phi + c3\phi) + \sqrt{2} c^3\phi^\prime (c\phi + c3\phi )\right] \nonumber \\
I_2 &=& - \frac{2s^2\phi}{9} \left[ 2 s\sigma c\sigma ( -2 \sqrt{2} c\sigma^\prime s\phi^\prime c\phi^\prime c\phi + c2\sigma^\prime s^2\phi^\prime + 2 c^2\phi^\prime c^2\phi ) \right.\nonumber \\ 
&& ~~~ \left. + c2\sigma ( s2\sigma^\prime s^2\phi^\prime - \sqrt{2} s\sigma^\prime s2\phi^\prime c\phi ) \right] ~.
\end{eqnarray}
The allowed regions of the $CPV$ phases for the combination $U_{13}(\phi,\sigma) U_{12}(\phi',\sigma')$ are shown in Fig. \ref{fig:13-12-libres}. In this case, given the relations of the correction parameters with the mixing angles (see Eq. \ref{eq:mixings1312}), it is not possible to restrict the $CP$ phases by means of a mixing angle, such that they remain free. However, by including the restriction of small deviations in the mass matrix, it is possible to restrict the allowed values of such phases in terms of the atmospheric angle. Despite the marked region of the phases obtained by including this last restriction, such predictions are in tension with recent suggestions of a Dirac phase different from zero \cite{deSalas:2017kay} and could be ruled out with forthcoming results. 

\begin{figure}\centering
\includegraphics[scale=0.4]{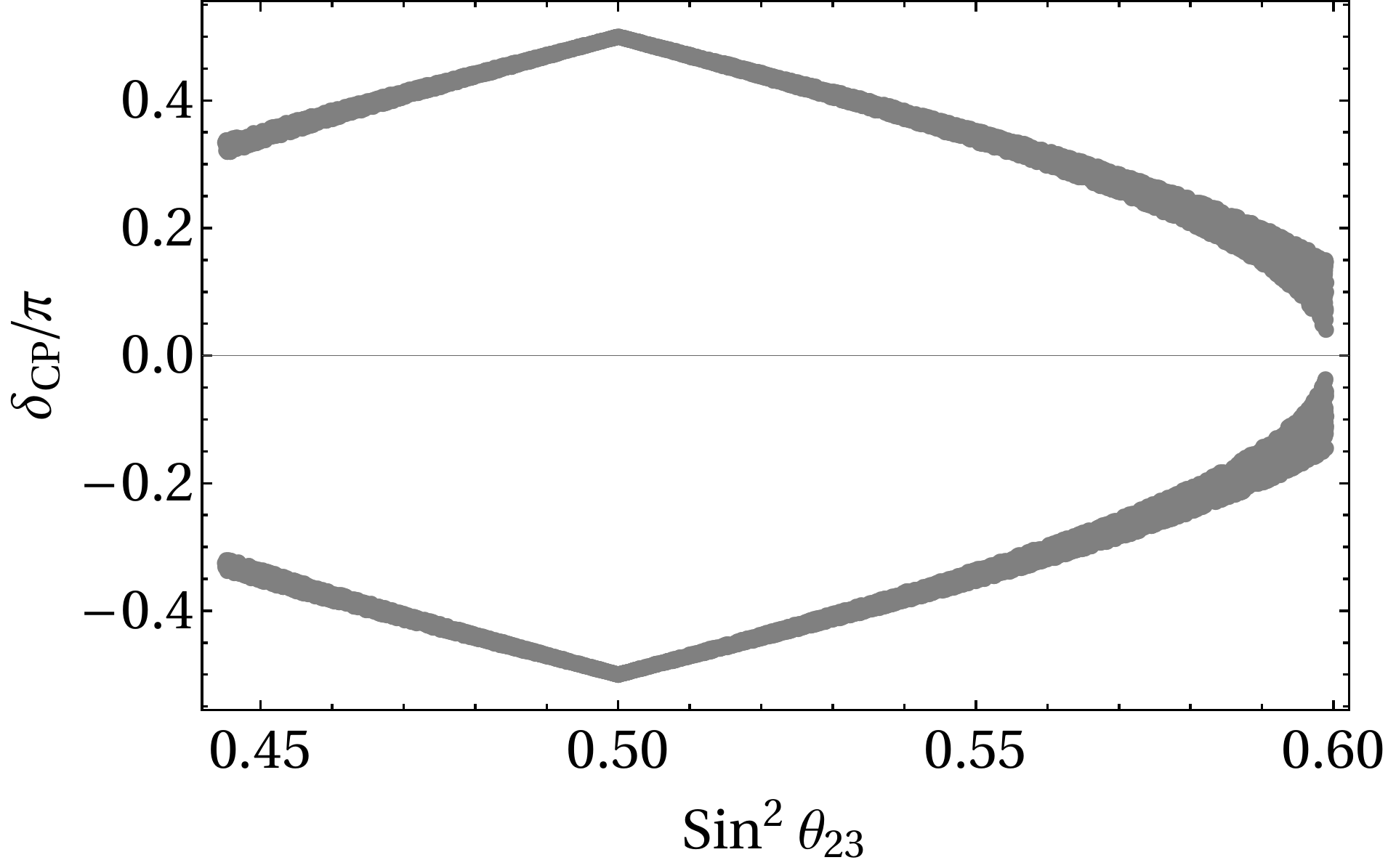}
\includegraphics[scale=0.4]{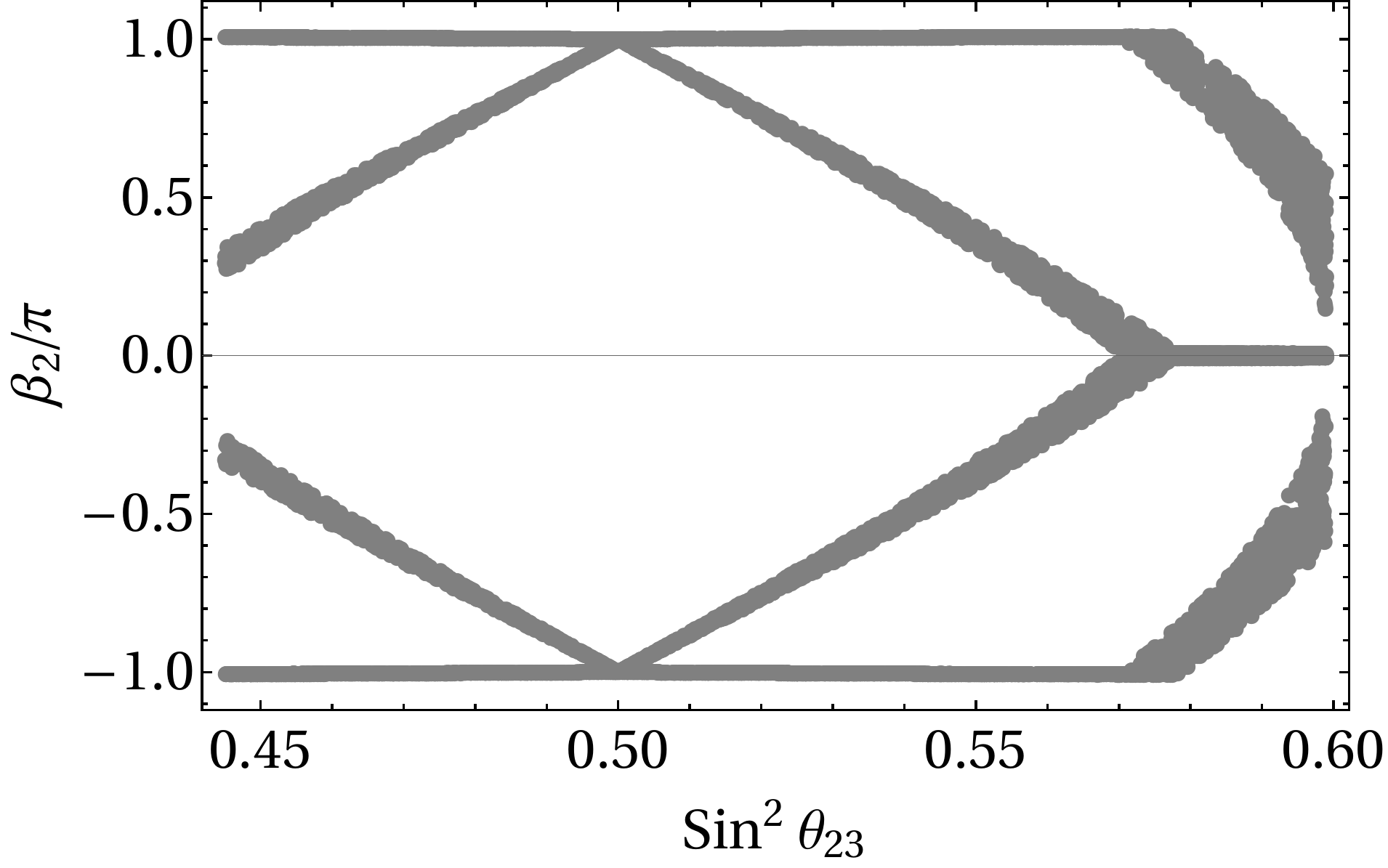}
\caption{Same description as in Fig. \ref{fig:12-13-libres} for the $\sigma = 0$.}\label{fig:13-12-cerolibre}
\end{figure}

In the specific case $\sigma=0$, we observe in Fig. \ref{fig:13-12-cerolibre} that $\delta_{CP}$ and $\beta_2$ can be related to the atmospheric angle. These phases can take values different from zero in a window near the central value, $\sin^2 \theta_{23} \sim 0.5 $, which is supported by global analysis; but they are in favor of small (or null) phases for values of an atmospheric angle at the end of its $3\sigma$ range. On the other hand, given the dependence of $I_1$ on the phase $\sigma'$, which is no further restricted by the solar angle relation of Eq. (\ref{eq:mixings1312}), the phase $\beta_1$ remains free and cannot be bounded in the present case. This case is not compatible with the requirement of small deviations in the mass matrix.
 
The $\sigma'=0$ does not give allowed regions for mixings and $CP$ phases. This can be easily verified by noting that the solar angle obtained from Eq. (\ref{eq:mixings1312}), $\sin^2 \theta_{12} \sim \frac{1}{2}(1+\sin^2 \phi')$, is always out of its allowed range for any given value of $\phi'$ in both mass orderings.

\begin{figure}\centering
\includegraphics[scale=0.6]{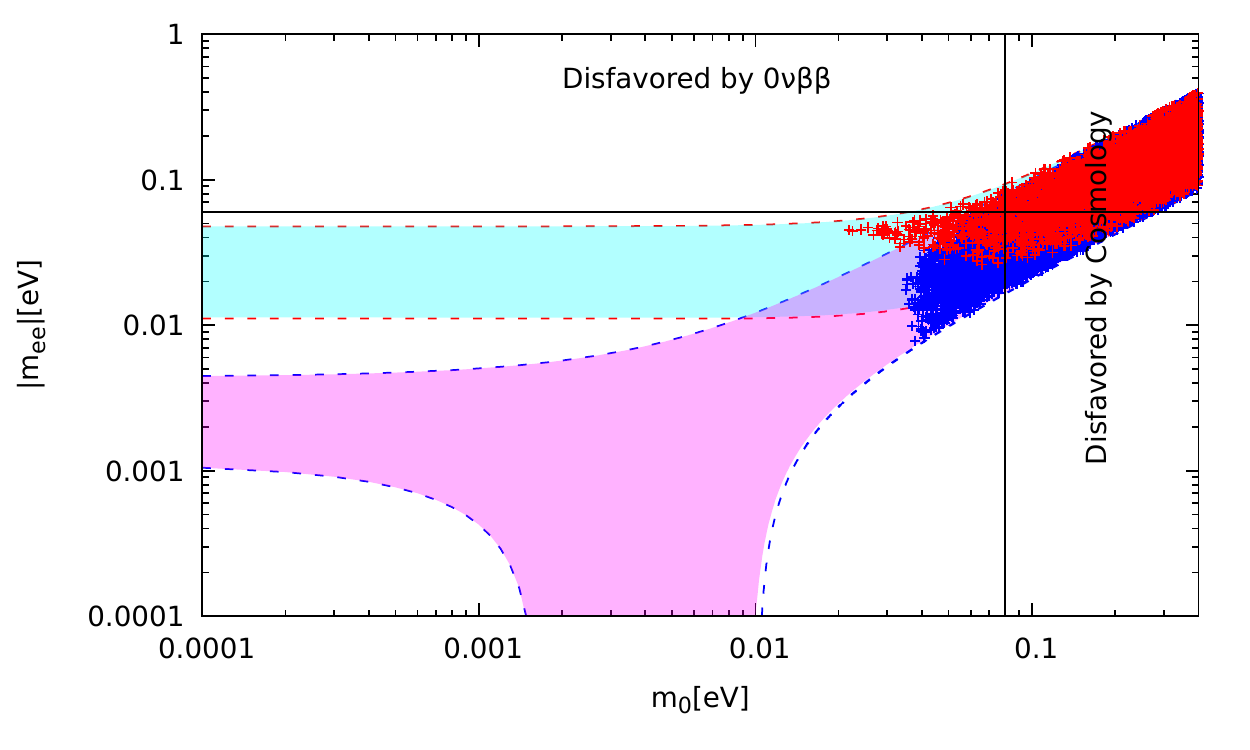} 
\includegraphics[scale=0.6]{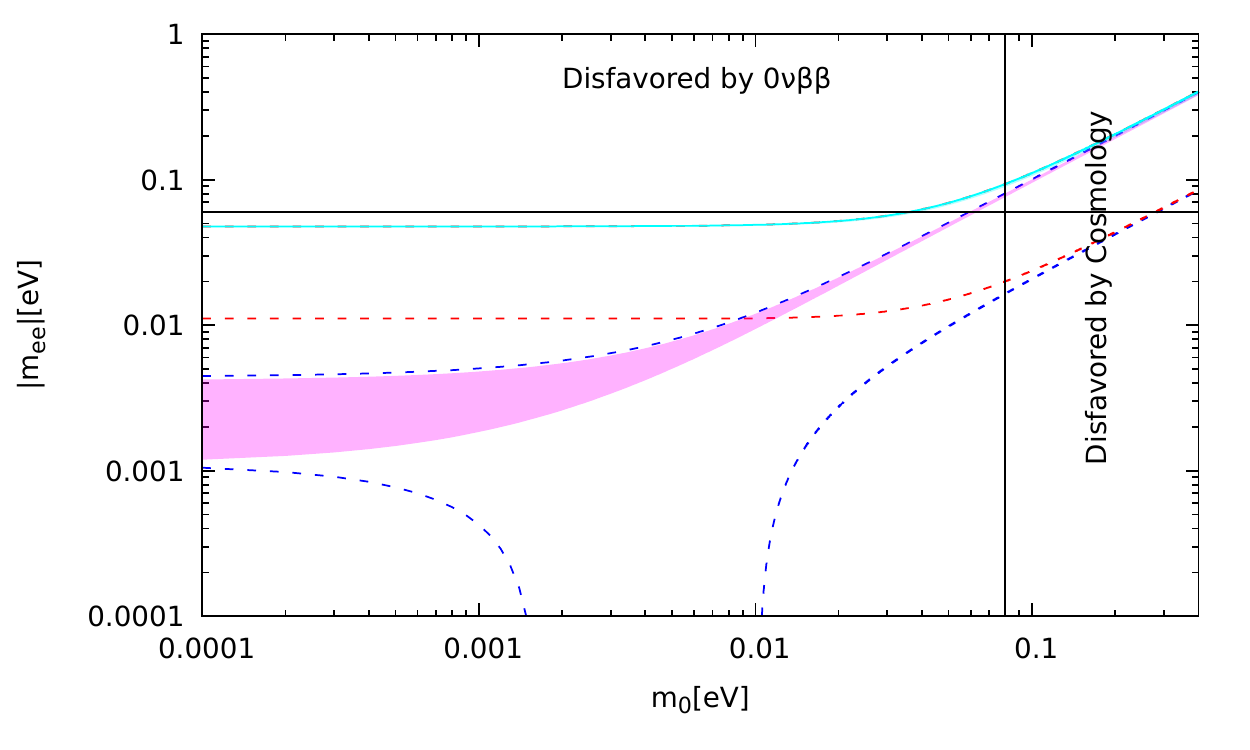}
\caption{Same description as in Fig. \ref{fig:12-13-0bb} but for $U_{13}(\phi,\sigma) U_{12}(\phi',\sigma')$ ~(left) and $\sigma=0$ ~(right).}\label{fig:13-12-0bb}
\end{figure}

The predicted values of $m_{ee}$ are shown in Fig. \ref{fig:13-12-0bb}. Given that the $CP$ phases cannot be bounded in the $U_{13}(\phi,\sigma) U_{12}(\phi',\sigma')$ case, $m_{ee}$ runs over the full available regions of the mass hierarchies. However, these regions are limited to the quasi-degenerate regime when the restriction of small deviations in the mass matrix is included. In the case of $\sigma=0$ , despite $\beta_1$ remains unbounded, well-defined regions are obtained for $m_{ee}$ given the sharpness of $\delta$ and $\beta_{2}$ in Fig. \ref{fig:13-12-cerolibre}. 

%%%%%%%%%%%%%%%%%%%%%%%%%%%%%%%%%%%%%%%%%%%%%%%%%%%%%%%%%%%%%%%%%%%%%%%%%%%%%%%%%%%%%%%%%%%%%%%%%%%%%%%%%%%%%%%%%%%%%%%%%%%%%%%
%%%%%%%%%%%%%%%%%%%%%%%%%%%%%%%%%%%%%%%%%%%%%%%%%%%%%%%%%%%%%%%%%%%%%%%%%%%%%%%%%%%%%%%%%%%%%%%%%%%%%%%%%%%%%%%%%%%%%%%%%%%%

\subsection{Case $U_{13}U_{23}$}\label{sec:1323}
Let us now consider the $U_{13}(\phi,\sigma)U_{23}(\phi',\sigma')$ case. The expressions relating the mixing angles with the correction parameters are given in this case by
\begin{eqnarray}\label{eq:mixings1323}
s^2 \theta_{12} &=& \frac{\sqrt{2} s2\phi^\prime s\phi c(\sigma - \sigma^\prime) + s^2\phi^\prime c2\phi - 1}{\sqrt{2} s2\phi^\prime s\phi c(\sigma-\sigma^\prime)-c^2\phi^\prime c2 \phi - 2} \nonumber \\
s^2 \theta_{13} &=& \frac{1}{3} \left[\sqrt{2} s2\phi^\prime s\phi c(\sigma-\sigma^\prime)-c^2\phi^\prime c2\phi + 1\right] \nonumber \\
s^2 \theta_{23} &=& \frac{1}{2} \left(\frac{\sqrt{3} c\sigma c^2\phi^\prime s2\phi - \sqrt{6} c\sigma^\prime s2\phi^\prime c\phi}{-\sqrt{2} s2\phi^\prime s\phi c(\sigma-\sigma^\prime) + c^2\phi^\prime c2\phi + 2}+1\right) ~.
\end{eqnarray}
The $CP$ invariants can be expressed as
\begin{eqnarray}\label{eq:invariants1323}
J_{CP} &=& \frac{c\phi}{24 \sqrt{3}} \nonumber \\
&& \times \left[\sqrt{2} s2\phi^\prime \left(-2 s^2\phi s(2 \sigma-\sigma^\prime)-3 s\sigma^\prime c2\phi + s\sigma^\prime \right)-8 s\sigma c2\phi^\prime s\phi \right] \nonumber \\
I_1 &=& \frac{4}{9} s\phi^\prime s\phi c^2\phi s(\sigma-\sigma^\prime) \left[ \sqrt{2} c\phi^\prime -2 s\phi^\prime s\phi c(\sigma-\sigma^\prime) \right] \nonumber \\
I_2 &=& -\frac{4 c^2\phi}{9}  \left[ \frac{s2\phi^\prime s\phi s(\sigma+\sigma^\prime)}{\sqrt{2}} + s2\sigma c^2\phi^\prime s^2\phi + s\sigma^\prime c\sigma^\prime s^2\phi^\prime\right] ~.
\end{eqnarray}
From Eqs. (\ref{eq:mixings1323}) and (\ref{eq:invariants1323}) we observed that the present combination leads to a set of relations in which each mixing and $CP$ invariant is related to the four correction parameters contrary to previous cases where it was possible to link, almost directly, one of the correction angles with one experimental angle. As a consequence, in the general case (totally free parameters) of this combination, no bounded regions for the predicted $CP$ phases were obtained, hence the plots are not shown.

\begin{figure}\centering
\includegraphics[scale=0.4]{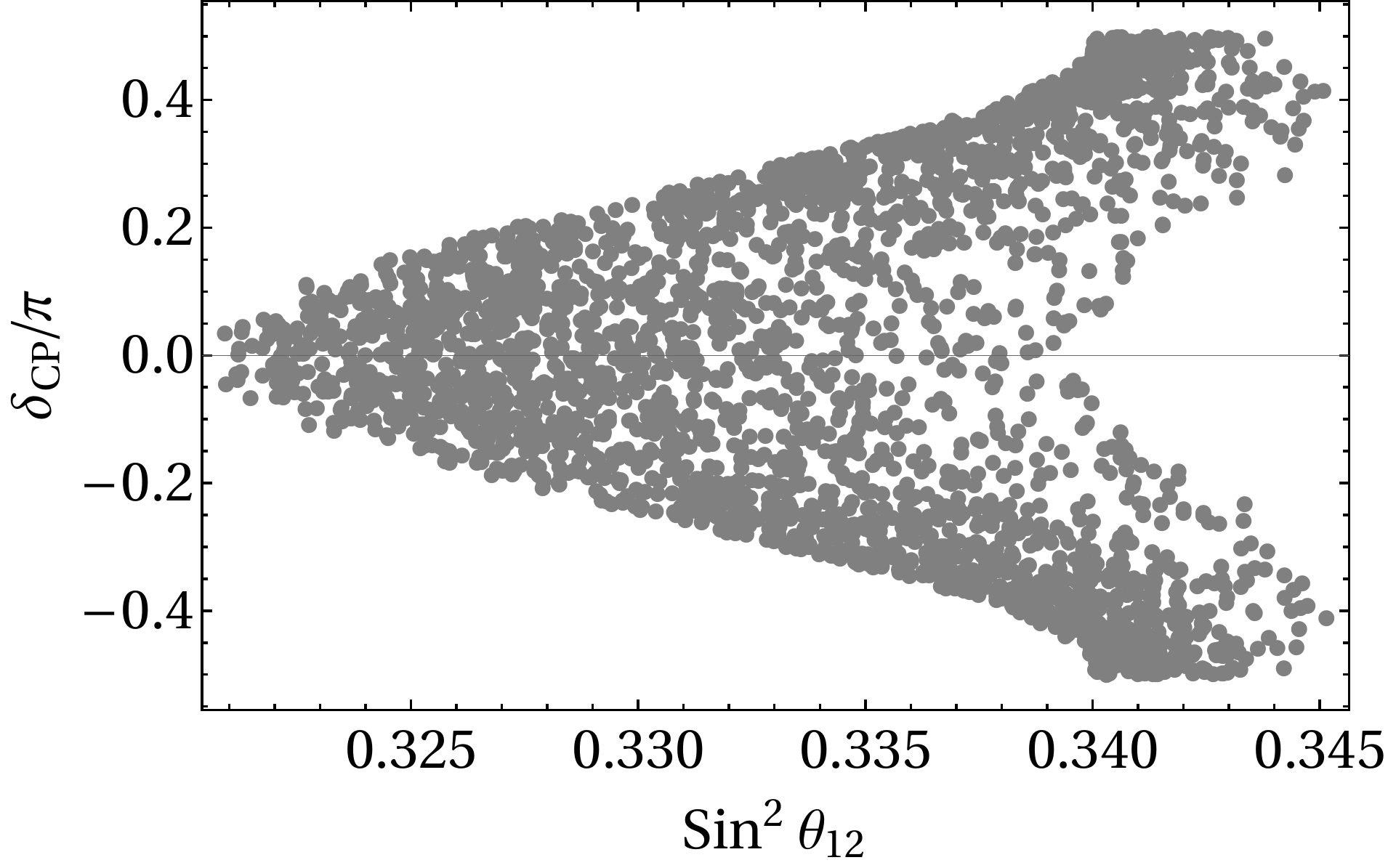}
\includegraphics[scale=0.4]{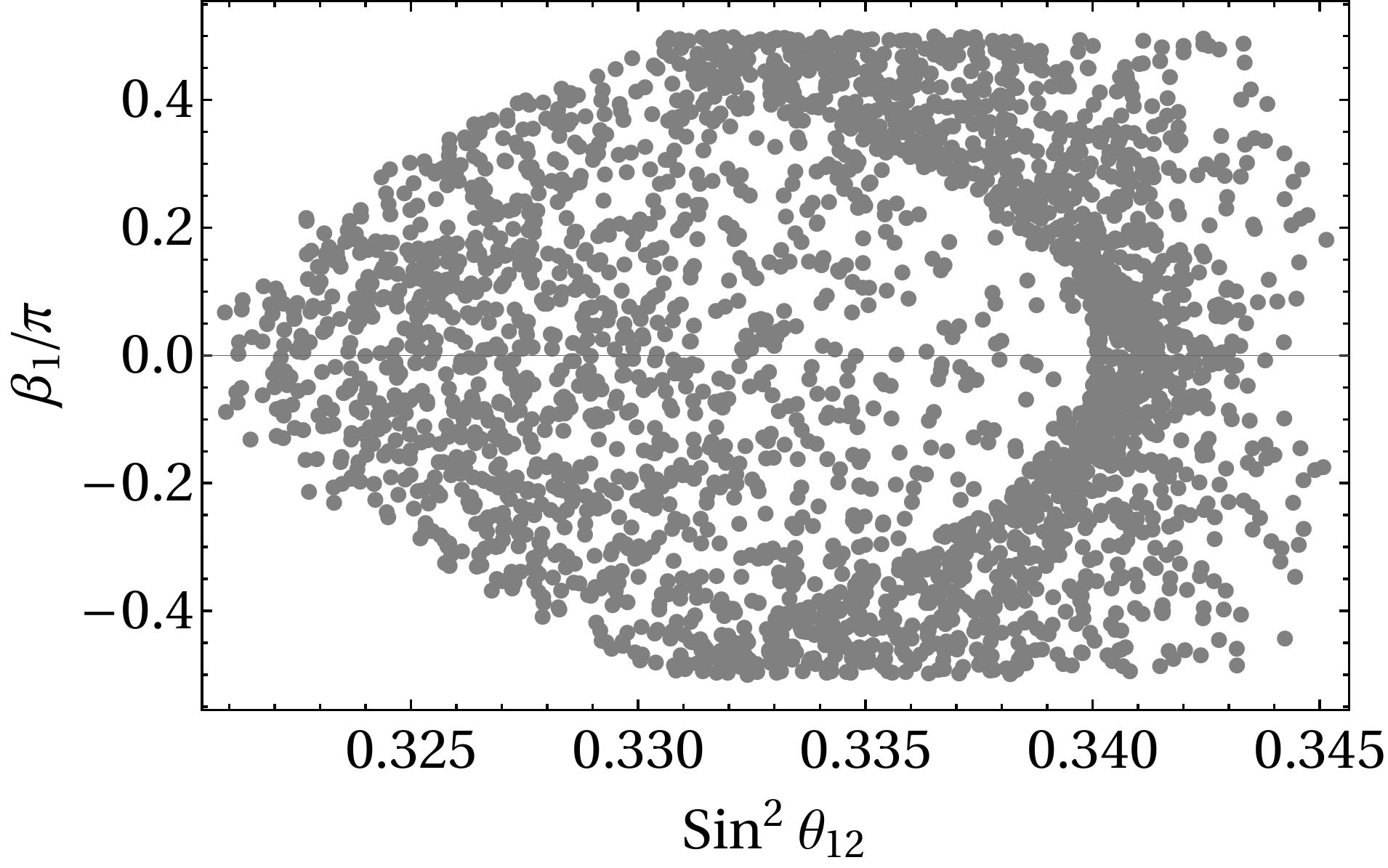}
\includegraphics[scale=0.4]{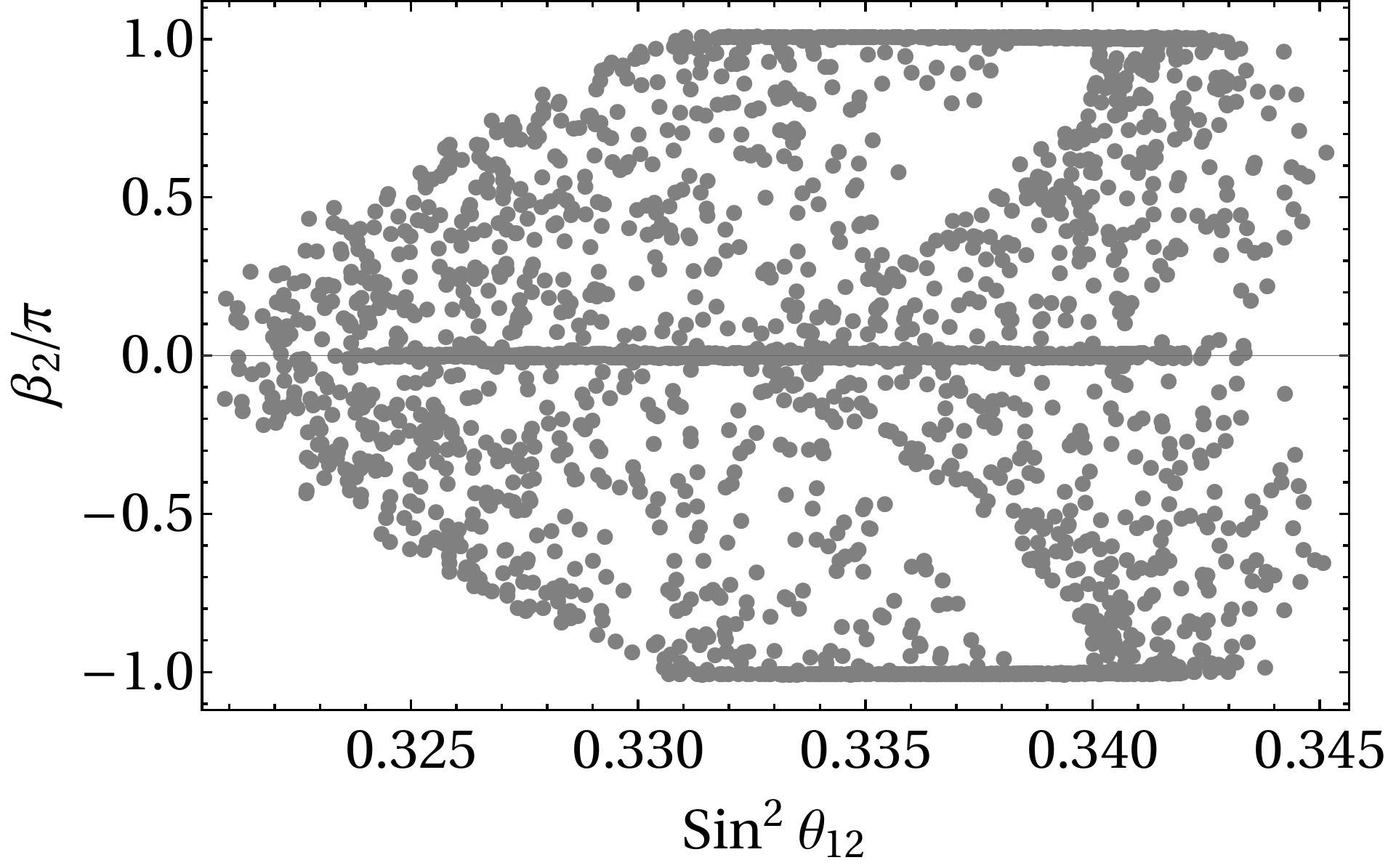}
\caption{Same description as in Fig. \ref{fig:12-13-libres} but for $\sigma' = 0$.}\label{fig:13-23-librecero}
\end{figure}

\begin{figure}\centering
\includegraphics[scale=0.4]{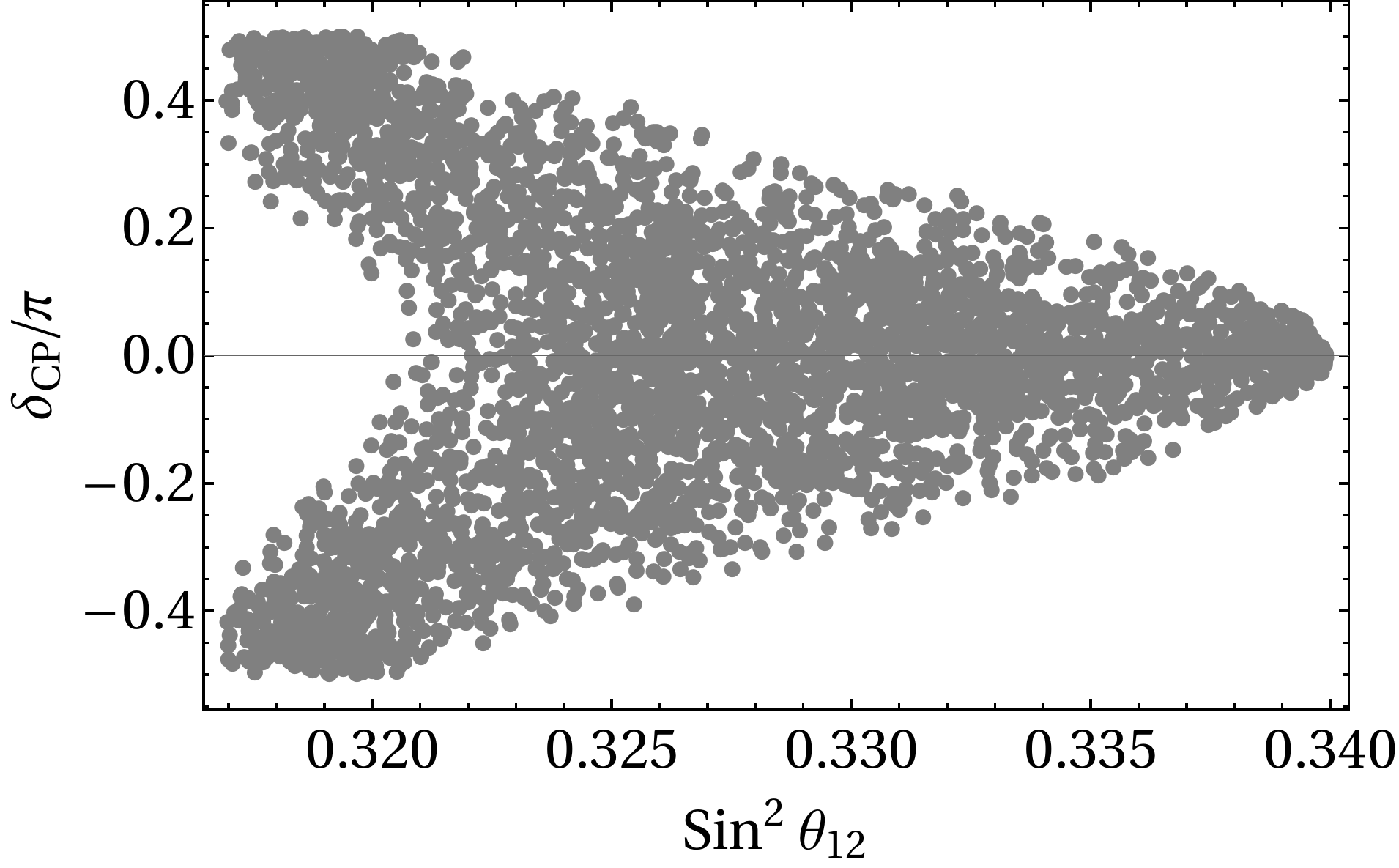}
\includegraphics[scale=0.4]{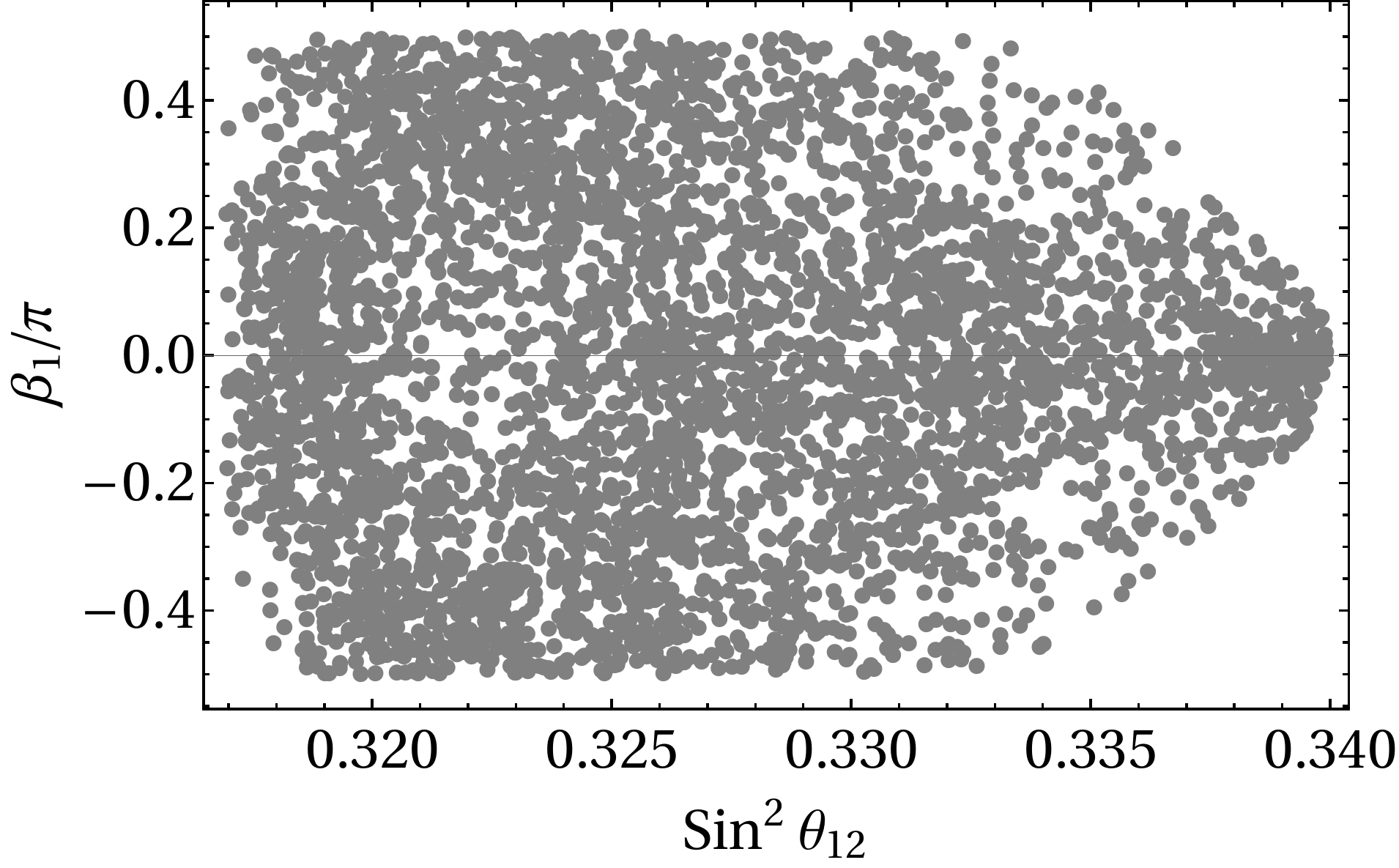}
\includegraphics[scale=0.4]{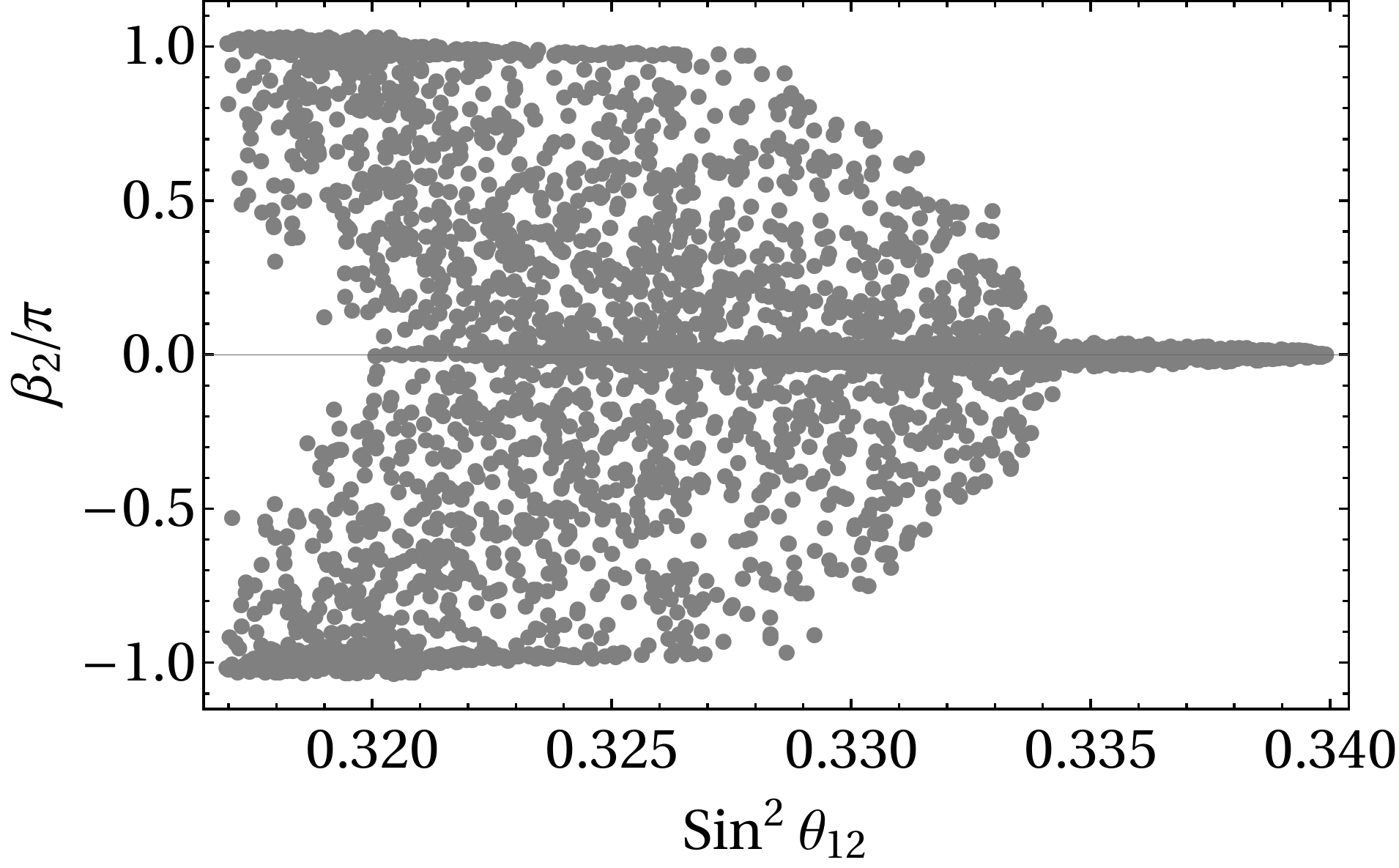}
\caption{Same description as in Fig. \ref{fig:12-13-libres} for the case $\sigma = 0$.}\label{fig:13-23-cerolibre}
\end{figure}

 In Fig. \ref{fig:13-23-librecero} we show the bounded regions of the $CPV$ phases related to the $\theta_{12}$ angle for the selection $\sigma' = 0$. We can observed that values of $\sin^2 \theta_{12}\sim 0.3$ are in favour of null $CP$ phases, which would be in tension with the current determinations of $\delta_{CP}$, but for values of $\sin^2 \theta_{12}\gtrsim 0.33$ the compatibility is restored, although this would leave the Majorana phases undetermined.
 
 For the $\sigma = 0$ case, we show the regions of the $CPV$ phases in terms of the $\theta_{12}$ angle in Fig. \ref{fig:13-23-cerolibre}. Opposite to the previous considerations, values of the solar angle near its lower $3\sigma$ limit suggest values for $\delta_{CP}$ and $\beta_2$ different from zero with $\beta_1$ undetermined, but for values near the upper limit, all the $CP$ phases tend to small values.
 
  Although no major bounded regions were obtained for the Majorana phases in the present cases, more precise measurements of the solar angle would help discriminating some of these scenarios \textit{via} the Dirac phase. In addition, none of these cases was compatible with the requirement of small deviations in the mass matrix. 
  
\begin{figure}\centering
\includegraphics[scale=0.6]{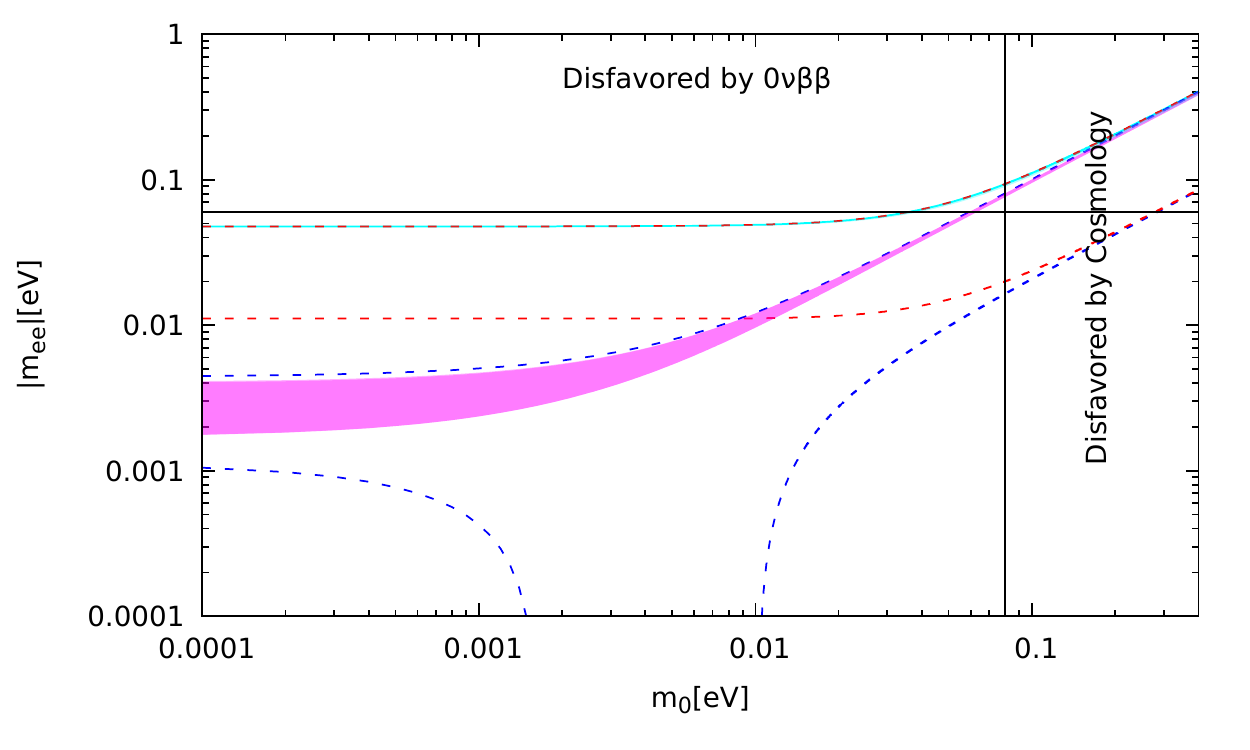} 
\caption{Same description as in Fig. \ref{fig:12-13-0bb} but for the cases $U_{13}(\phi,\sigma) U_{23}(\phi',\sigma')$, ~$\sigma = 0$, and $\sigma' = 0$.}\label{fig:13-23-0bb}
\end{figure}

In Fig. \ref{fig:13-23-0bb}, we show the predicted region of $m_{ee}$ for both mass hierarchies. No visible differences were found between the three analysed cases, $U_{13}(\phi,\sigma) U_{23}(\phi',\sigma')$, $\sigma=0$, and $\sigma' = 0$, but sharp regions were obtained. For reference, we show only one plot which is valid for the three cases.  
 
%%%%%%%%%%%%%%%%%%%%%%%%%%%%%%%%%%%%%%%%%%%%%%%%%%%%%%%%%%%%%%%%%%%%%%%%%%%%%%%%%%%%%%%%%%%%%%%%%%%%%%%%%%%%%%%%%%%%%%%%%%%%%%%
%%%%%%%%%%%%%%%%%%%%%%%%%%%%%%%%%%%%%%%%%%%%%%%%%%%%%%%%%%%%%%%%%%%%%%%%%%%%%%%%%%%%%%%%%%5%%%%%%%%%%%%%%%%%%%%%%%%%
\subsection{Case $U_{23}U_{12}$}
 The relations of the correction parameters in the case $U_{23}(\phi,\sigma)U_{12}(\phi',\sigma')$ are given by
\begin{eqnarray}\label{eq:mixings2312}
s^2 \theta_{12} &=& -\frac{\sqrt{2} c\sigma^\prime c\phi s2\phi^\prime + c^2\phi c^2\phi^\prime + 2 s^2\phi^\prime}{s^2\phi - 3} \nonumber \\
s^2 \theta_{13} &=& \frac{s^2\phi}{3} \nonumber \\
s^2 \theta_{23} &=& \frac{1}{2}-\frac{\sqrt{6} c\sigma s2\phi}{c2\phi + 5} ~,
\end{eqnarray}
and the $CP$ invariants are
\begin{eqnarray}\label{eq:invariants2312}
J_{CP} &=& \frac{1}{48 \sqrt{3}} \nonumber \\
&&\times \left[ s2\phi^\prime \left( c\sigma^\prime s\sigma ( s3\phi - 7 s\phi ) + s\sigma^\prime c\sigma (5 s3\phi - 3 s\phi ) \right) - 4 \sqrt{2} s\sigma s2\phi c2\phi^\prime  \right] \nonumber \\
I_1 &=& \frac{1}{9} s\sigma^\prime s\phi^\prime \nonumber \\
&&\times \left[ -8 c\sigma^\prime c2\sigma^\prime c^2\phi s^3\phi^\prime - 2 c\sigma^\prime s\phi^\prime c^2\phi^\prime \left(c^4\phi - 8 c^2\phi + 4\right) \right. \nonumber \\
&& ~~~ \left. - \frac{5 c\phi - c3\phi}{2 \sqrt{2}} \left(-4 c2\sigma^\prime s^2\phi^\prime c\phi^\prime + c\phi^\prime + c3\phi^\prime \right) \right] \nonumber \\
I_2 &=& -\frac{s^2\phi}{9} \left[ c\phi \left(c\phi s^2\phi^\prime s2(\sigma^\prime+\sigma)-\sqrt{2} s2\phi^\prime s(\sigma^\prime+2 \sigma)\right)+2 s2\sigma c^2\phi^\prime \right] ~.
\end{eqnarray}
From Eq. (\ref{eq:mixings2312}) we observe that the $\phi$ angle is directly restricted by the reactor angle and that restrictions on $\sigma^{(\prime)}$ are only due to the atmospheric (solar) angle. In the general case, $U_{23}(\phi,\sigma)U_{12}(\phi',\sigma')$, the experimental mixing angles are fully reproduced by the correction parameters without visible restrictions on the $CP$ phases, such that the plots are not shown. 

\begin{figure}\centering
\includegraphics[scale=0.4]{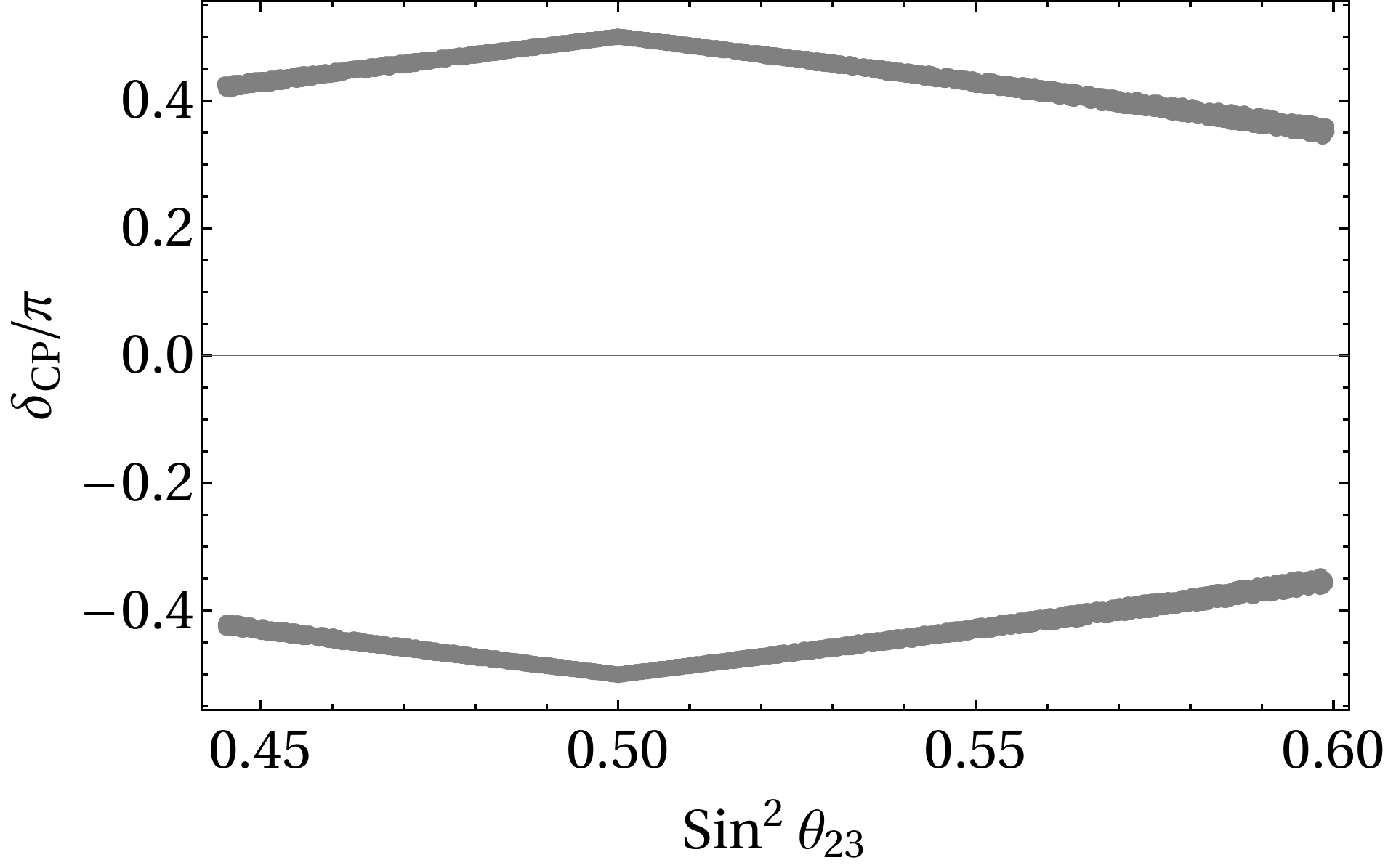}
\includegraphics[scale=0.4]{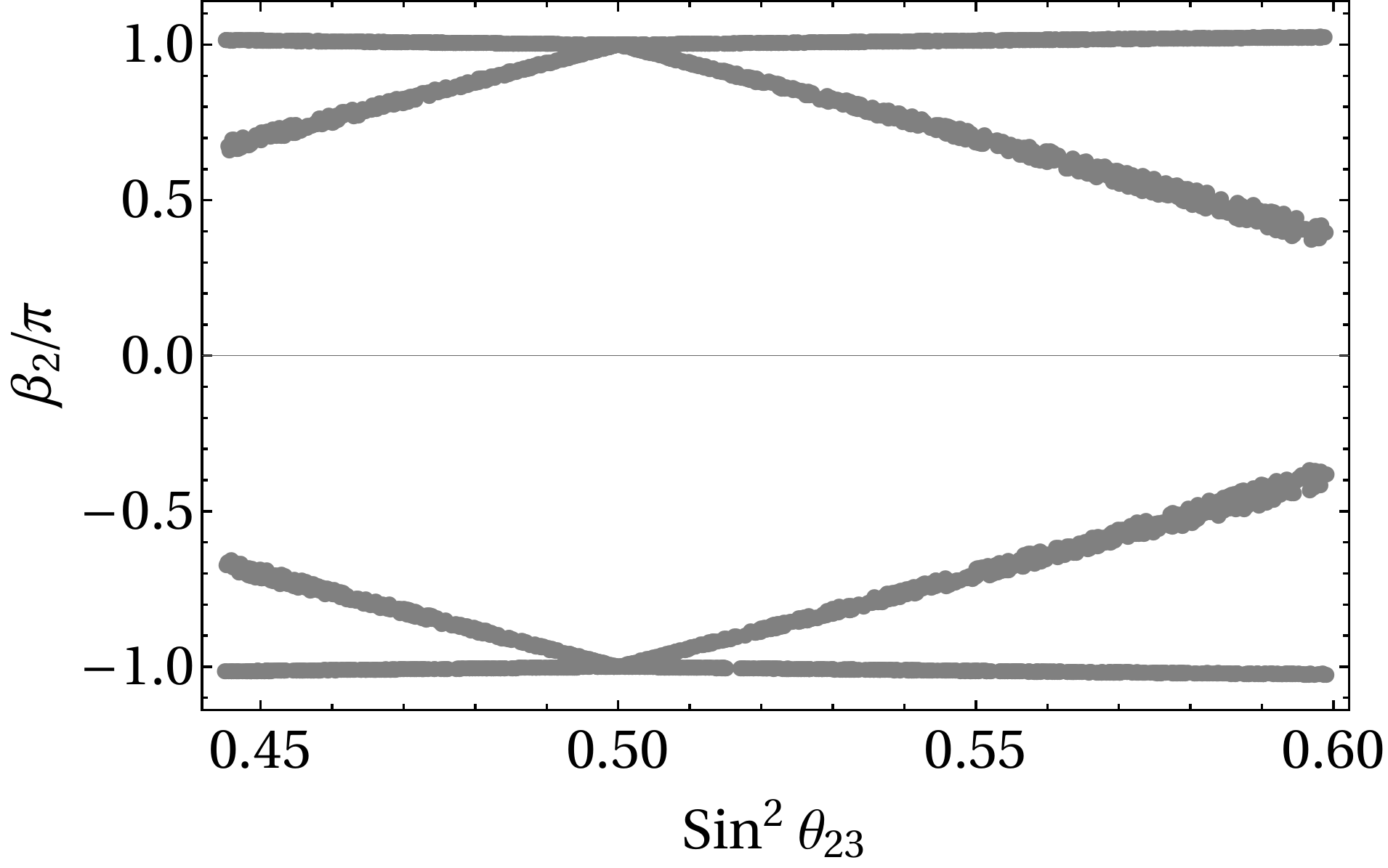}
\caption{Same description as in Fig. \ref{fig:12-13-libres} for $\sigma' = 0$.}\label{fig:23-12-cerolibre}
\end{figure}

For $\sigma=0$, no regions were obtained. This can be verified by noting that in this case $\phi\sim 0.25$, given the restrictions coming from the reactor angle in Eq. (\ref{eq:mixings2312}), which always leads to a value of the atmospheric angle out of the $3\sigma$ range. 

In the $\sigma' = 0$ selection, $\beta_1$ is fixed to zero given that $I_1 = 0$. The regions for $\delta_{CP}$ and $\beta_2$ in terms of the atmospheric angle are shown in Fig. \ref{fig:23-12-cerolibre}. The sharp regions obtained in this case suggest values of $\delta_{CP}$ and $\beta_2$ different from zero for the full range of the experimental mixing, which could be additionally bounded with most precise determinations of the atmospheric angle. No further restrictions can be imposed to the cases of the present section as they are not compatible with the requirement of small $\mu-\tau$ breaking in the mass matrix.

\begin{figure}\centering
\includegraphics[scale=0.6]{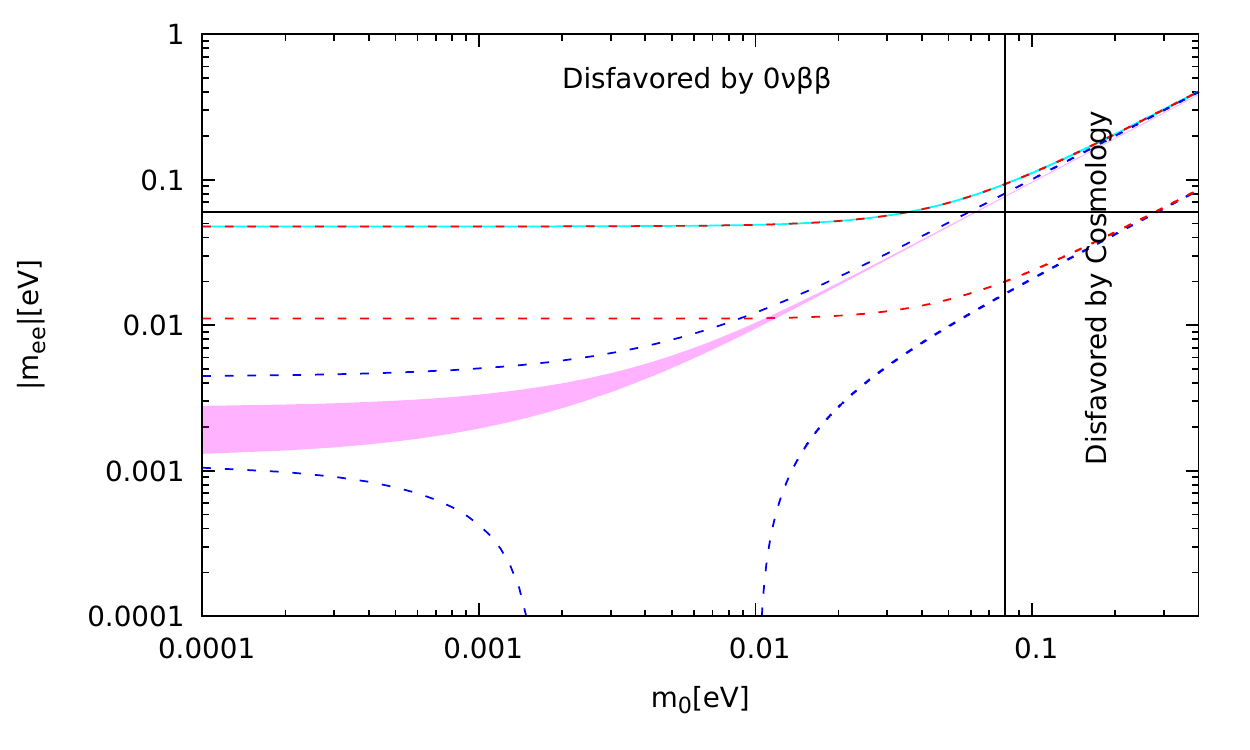} 
\caption{Same description as in Fig. \ref{fig:12-13-0bb} but for $\sigma'=0$.}\label{fig:23-12-0bb}
\end{figure}

In Fig. \ref{fig:23-12-0bb}, we show the predictions of $m_{ee}$ for the case $\sigma' = 0$. Sharp regions can be observed for both mass hierarchies, which is understood given the very limited values of the $CP$ phases in this combination. On the other hand, for the general case, $U_{23}(\phi,\sigma)U_{12}(\phi',\sigma)$, no differences were observed compared to the case of free $CP$ phases, hence the plot is not shown.

%%%%%%%%%%%%%%%%%%%%%%%%%%%%%%%%%%%%%%%%%%%%%%%%%%%%%%%%%%%%%%%%%%%%%%%%%%%%%%%%%%%%%%%%%%%%%%%%%%%%%%%%%%%%%%%%%%%%%%%%%%%%%%%
%%%%%%%%%%%%%%%%%%%%%%%%%%%%%%%%%%%%%%%%%%%%%%%%%%%%%%%%%%%%%%%%%%%%%%%%%%%%%%%%%%%%%%%%%%%%%%%%%%%%%%%%%%%%%%%%%

\subsection{Case $U_{23}U_{13}$}
\begin{figure}\centering
\includegraphics[scale=0.4]{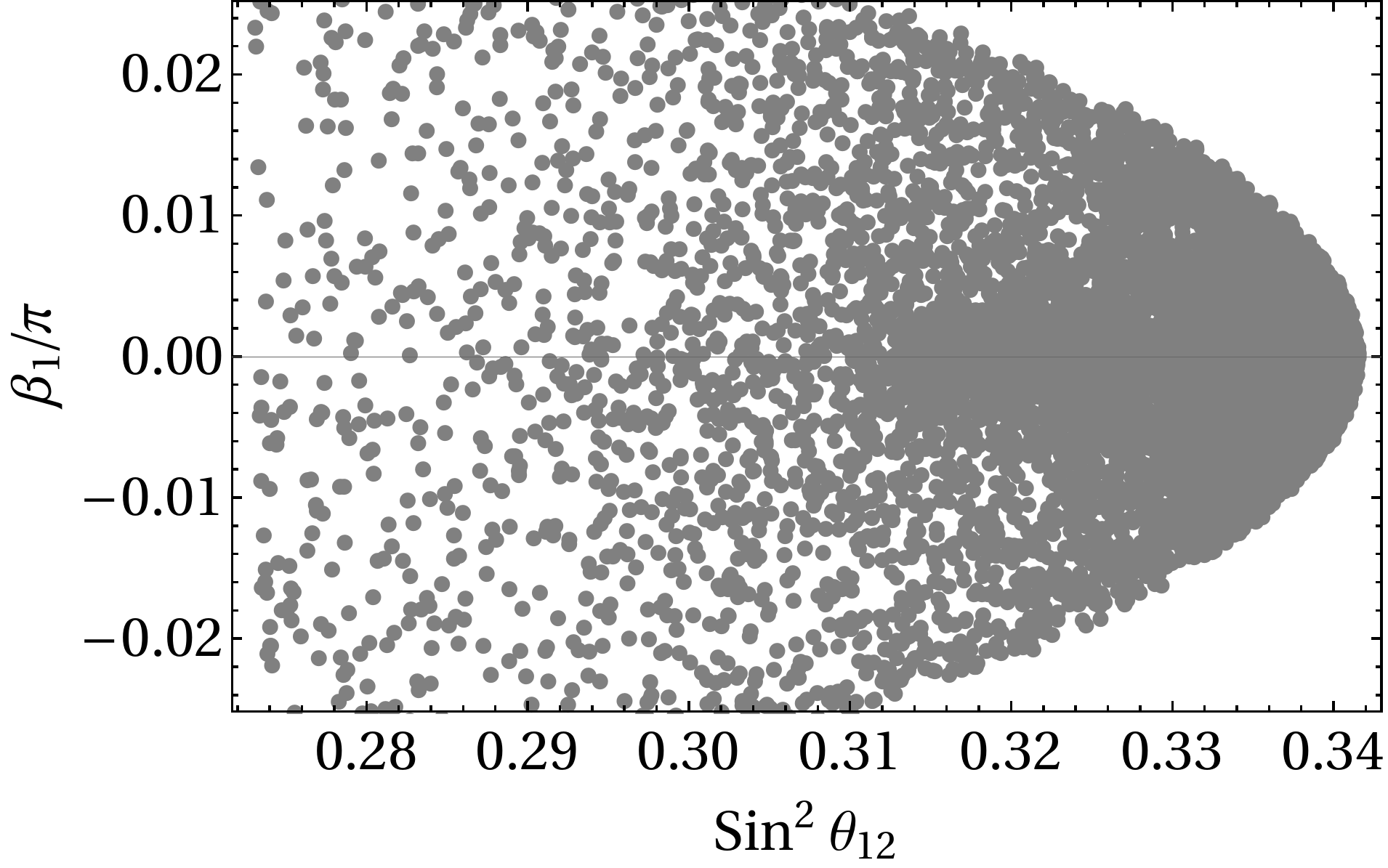}
\caption{Same description as in Fig. \ref{fig:12-13-libres} for the $U_{23}(\phi,\sigma)U_{13}(\phi',\sigma')$ case.}\label{fig:23-13-libres}
\end{figure}

 Our last case of interest is the $U_{23}(\phi,\sigma)U_{13}(\phi',\sigma')$ combination. The experimental angles can be related to the corrections parameters by
\begin{eqnarray}\label{eq:mixings2313}
s^2 \theta_{12} &=& \left[ \frac{1}{2} \sec ^2\phi \left(-2 \sqrt{2} s\phi s2\phi^\prime c(\sigma^\prime-\sigma)+c2\phi^\prime + 3\right)+c^2\phi^\prime \right]^{-1} \nonumber \\
s^2 \theta_{13} &=& \frac{1}{3} \left[\sqrt{2} s\phi s2\phi^\prime c(\sigma^\prime-\sigma)+ s^2\phi c^2\phi^\prime + 2 s^2\phi^\prime  \right] \nonumber \\
s^2 \theta_{23} &=& -3\left[ \sqrt{2} s\phi s2\phi^\prime c(\sigma^\prime-\sigma)+s^2\phi c^2\phi^\prime + 2s^2\phi^\prime - 3\right]^{-1} \nonumber \\
&& \left[ \left(\frac{c\sigma^\prime s\phi^\prime}{\sqrt{6}}-\frac{c\sigma s\phi c\phi^\prime}{\sqrt{3}}+\frac{c\phi c\phi^\prime}{\sqrt{2}}\right)^2+\left(\frac{s\sigma s\phi c\phi^\prime}{\sqrt{3}}-\frac{s\sigma^\prime s\phi^\prime}{\sqrt{6}}\right)^2\right] ~.
\end{eqnarray}
The $CP$ invariants which lead to the relations between the correction parameters and the $CPV$ phases are given in this case by
\begin{eqnarray}\label{eq:invariants2313}
J_{CP} &=& -\frac{1}{24 \sqrt{3}}\left[ c\phi s2\phi^\prime \left(-4 s^2\phi s(\sigma^\prime-2 \sigma)+3s\sigma^\prime c2\phi + s\sigma^\prime \right )+2 \sqrt{2} s\sigma s2\phi c2\phi^\prime \right] \nonumber \\
I_1 &=& \frac{2}{9} s\phi c^2\phi s\phi^\prime s(\sigma^\prime-\sigma) \left[\sqrt{2} c\phi^\prime - s\phi s\phi^\prime c(\sigma^\prime-\sigma) \right] \nonumber \\
I_2 &=& \frac{1}{36} \left[-8 s^2\phi s^4\phi^\prime s(4\sigma^\prime-2 \sigma) - 2 \sqrt{2} (5 s\phi + s3\phi) s\phi^\prime c^3\phi^\prime s(\sigma^\prime+\sigma) \right. \nonumber \\
&& +2 \sqrt{2} (5 s\phi + s3\phi) s^3\phi^\prime c\phi^\prime s(3 \sigma^\prime-\sigma) \nonumber \\
&&  \left. -\frac{1}{8} s2\sigma^\prime s^22\phi^\prime \left(28 c2\phi + c4\phi + 3 \right) -8 s2\sigma s^2\phi c^4\phi^\prime \right] ~.
\end{eqnarray}
As it was noted in the $U_{13}U_{23}$ case (see Sec. \ref{sec:1323}), no major restrictions can be imposed over the correction parameters as each mixing angle in Eq. (\ref{eq:mixings2313}) depends on the four correction parameters; which explains why no bounded regions for $\delta_{CP}$ and $\beta_2$ were obtained in the general (free parameters) case. On the other hand, because of the suppression by the constant factor and the phase cancellations in $I_1$ (see Eq. \ref{eq:invariants2313}), the Majorana phase $\beta_1$ is restricted to small values. In Fig. \ref{fig:23-13-libres} we show the allowed region of $\beta_1$ in terms of the solar angle. Furthermore, such combination is not compatible with perturbative deviations in the $\mu-\tau$ symmetric mass matrix such that no additional restrictions can be imposed on the $CP$ phases.

\begin{figure}\centering
\includegraphics[scale=0.4]{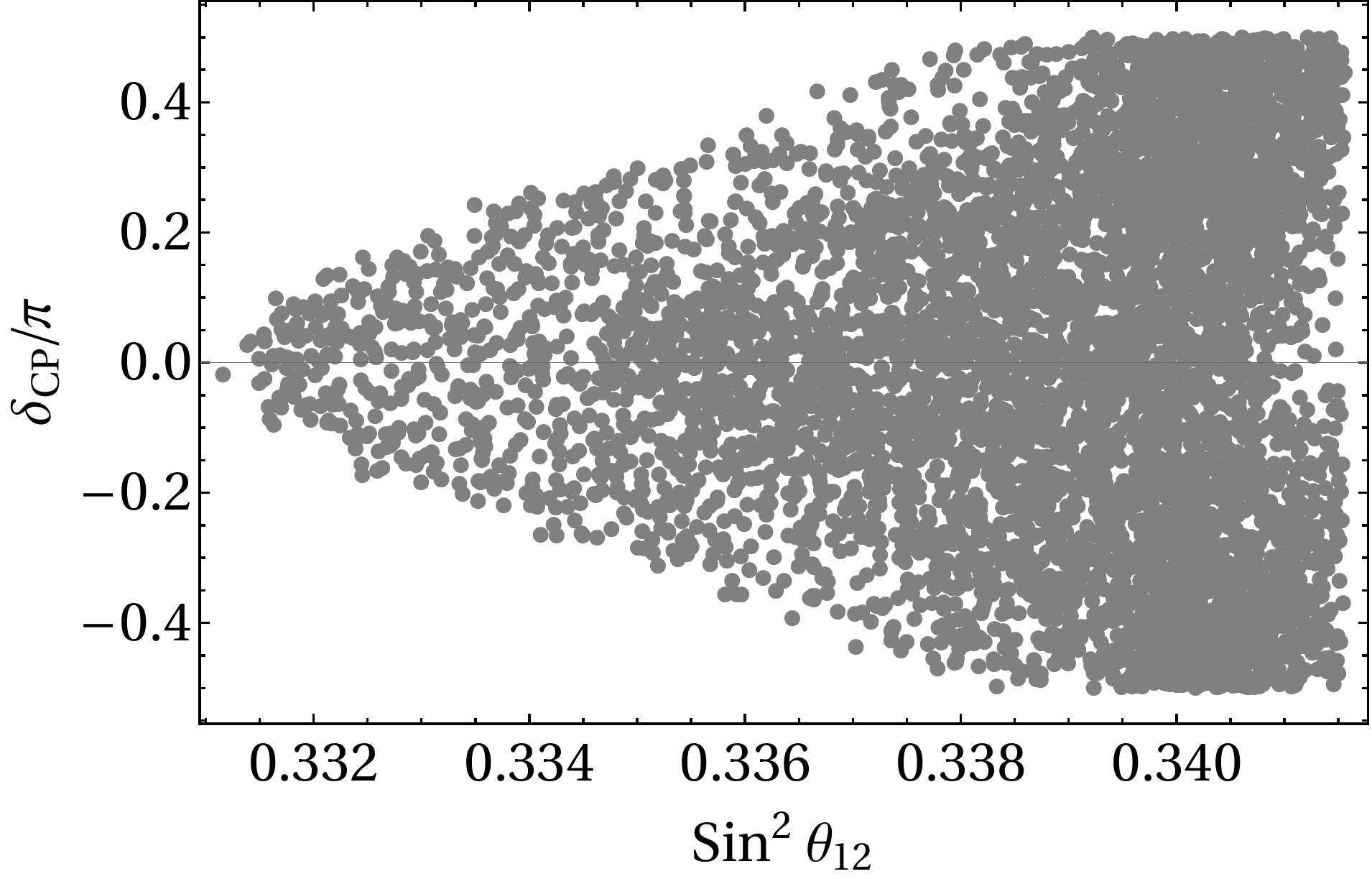}
\includegraphics[scale=0.4]{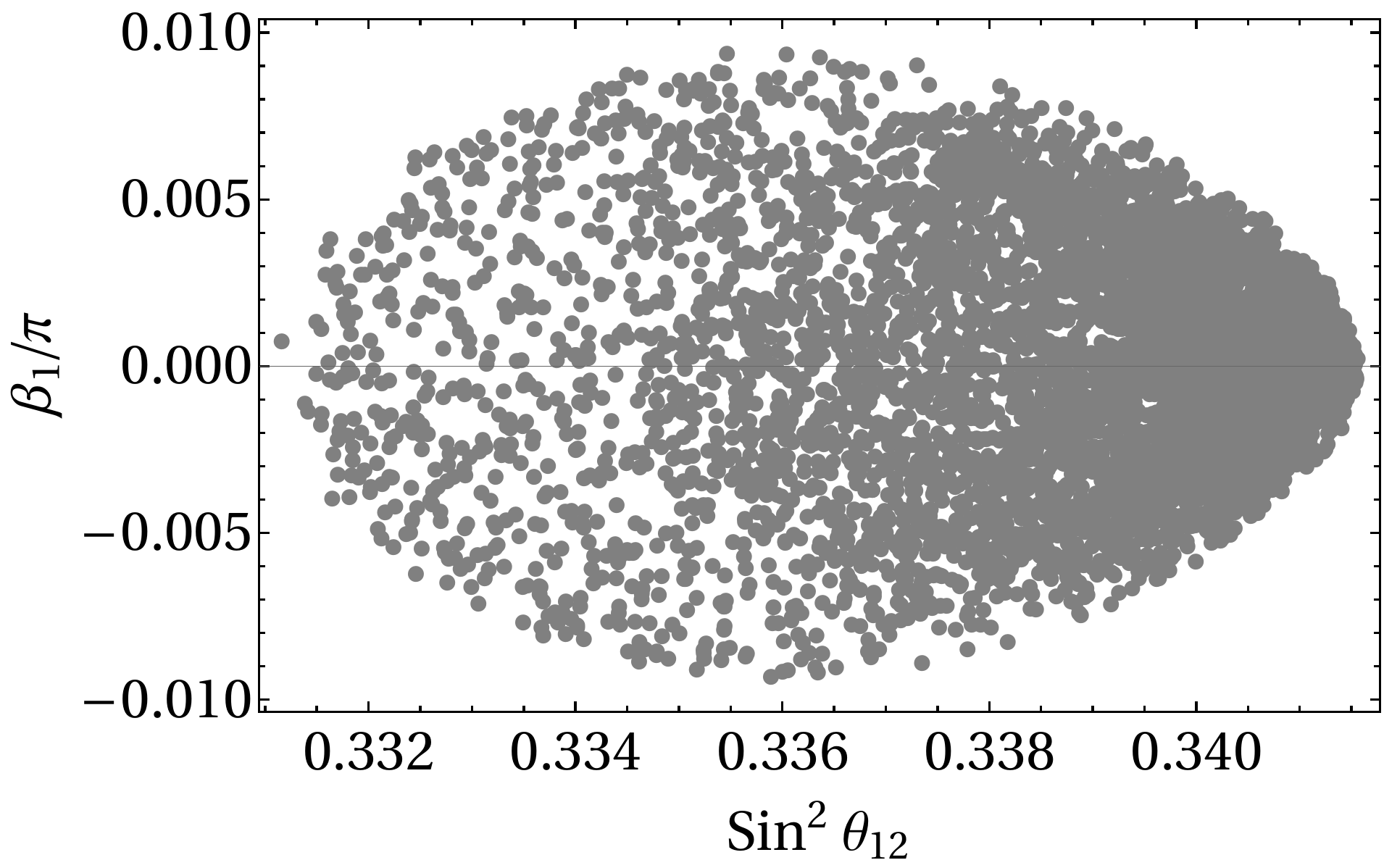}
\includegraphics[scale=0.4]{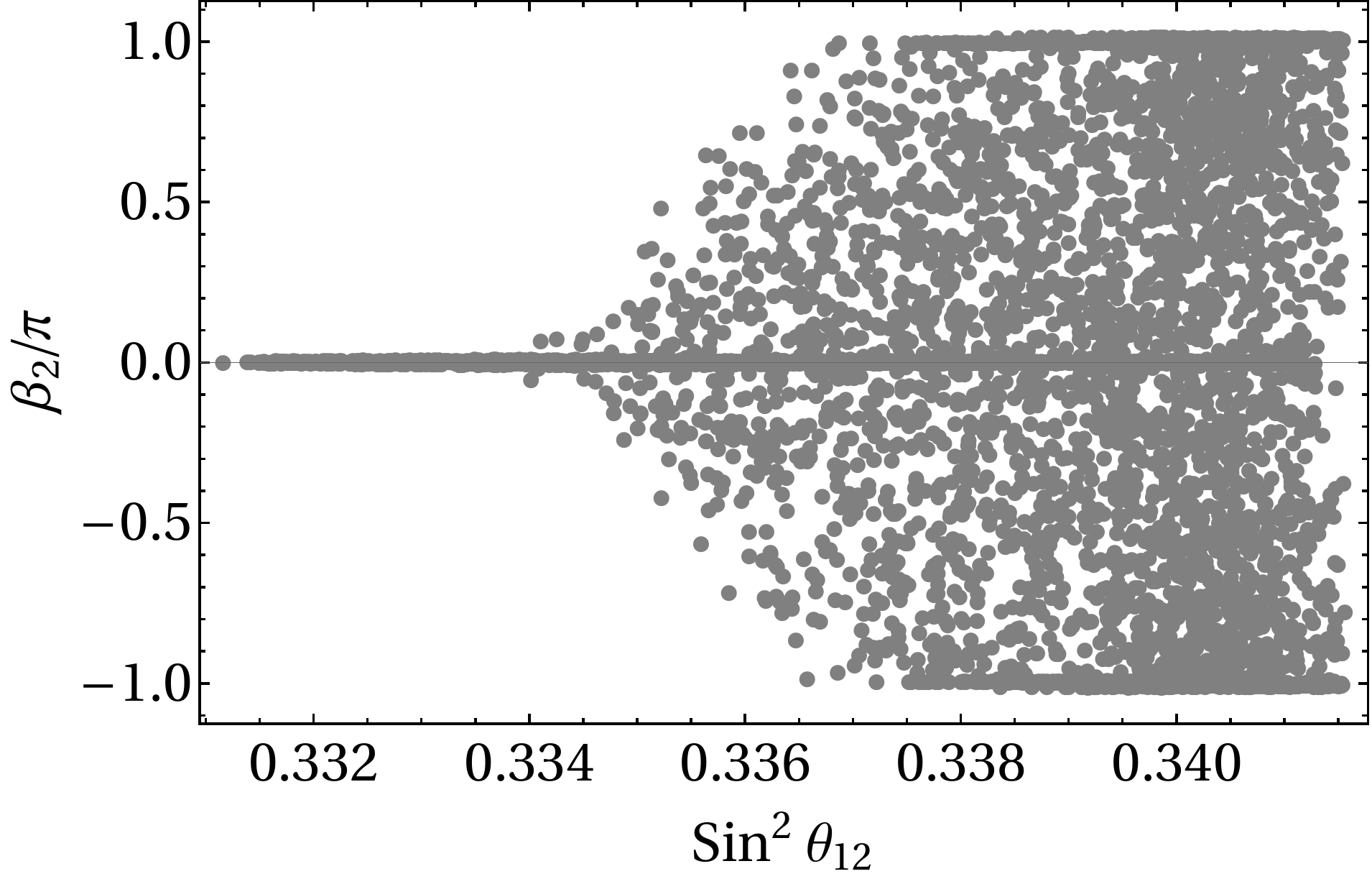}
\caption{Same description as in Fig. \ref{fig:12-13-libres} for $\sigma = 0$.}\label{fig:23-13-librecero}
\end{figure}

\begin{figure}\centering
\includegraphics[scale=0.4]{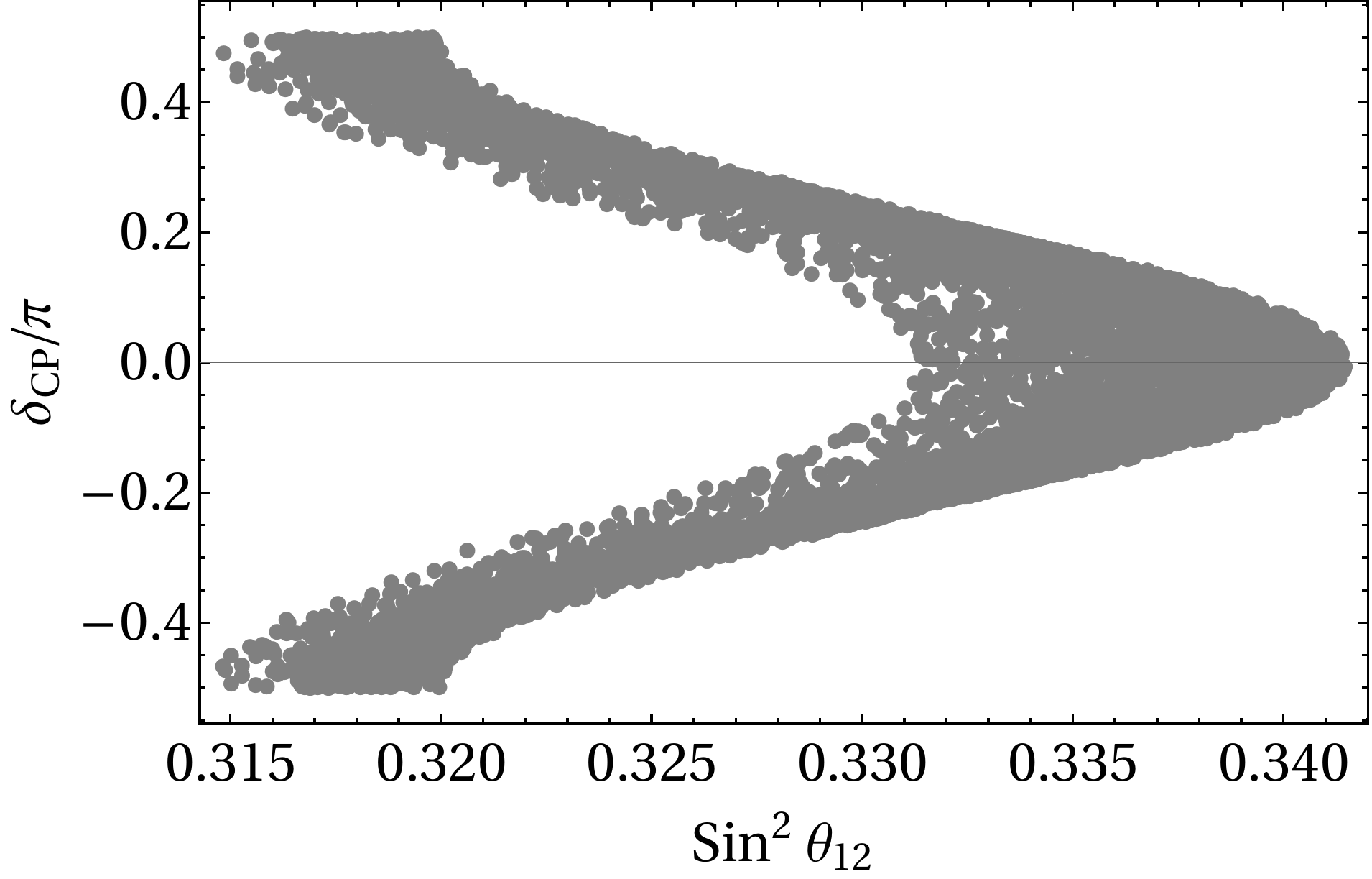}
\includegraphics[scale=0.4]{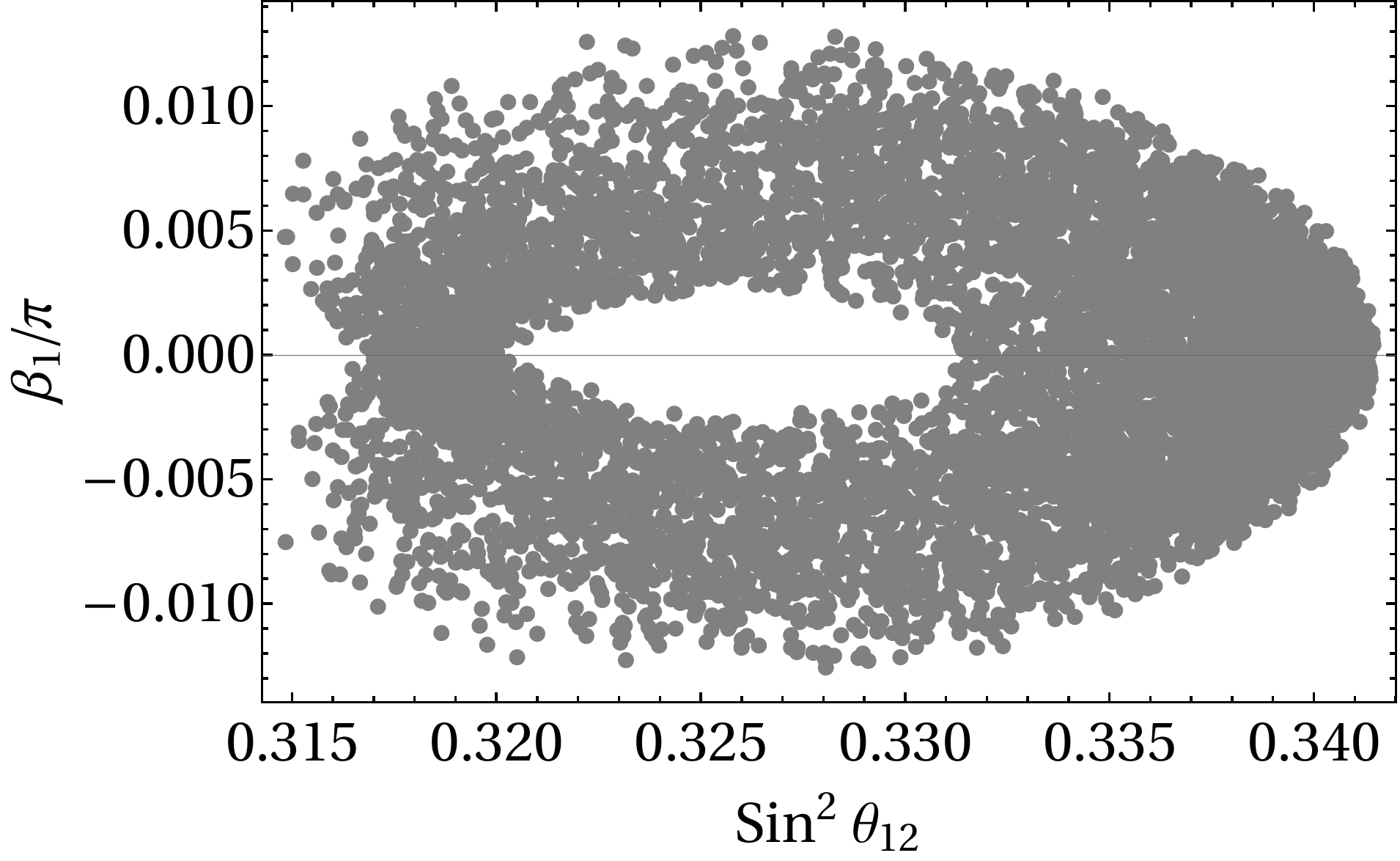}
\includegraphics[scale=0.4]{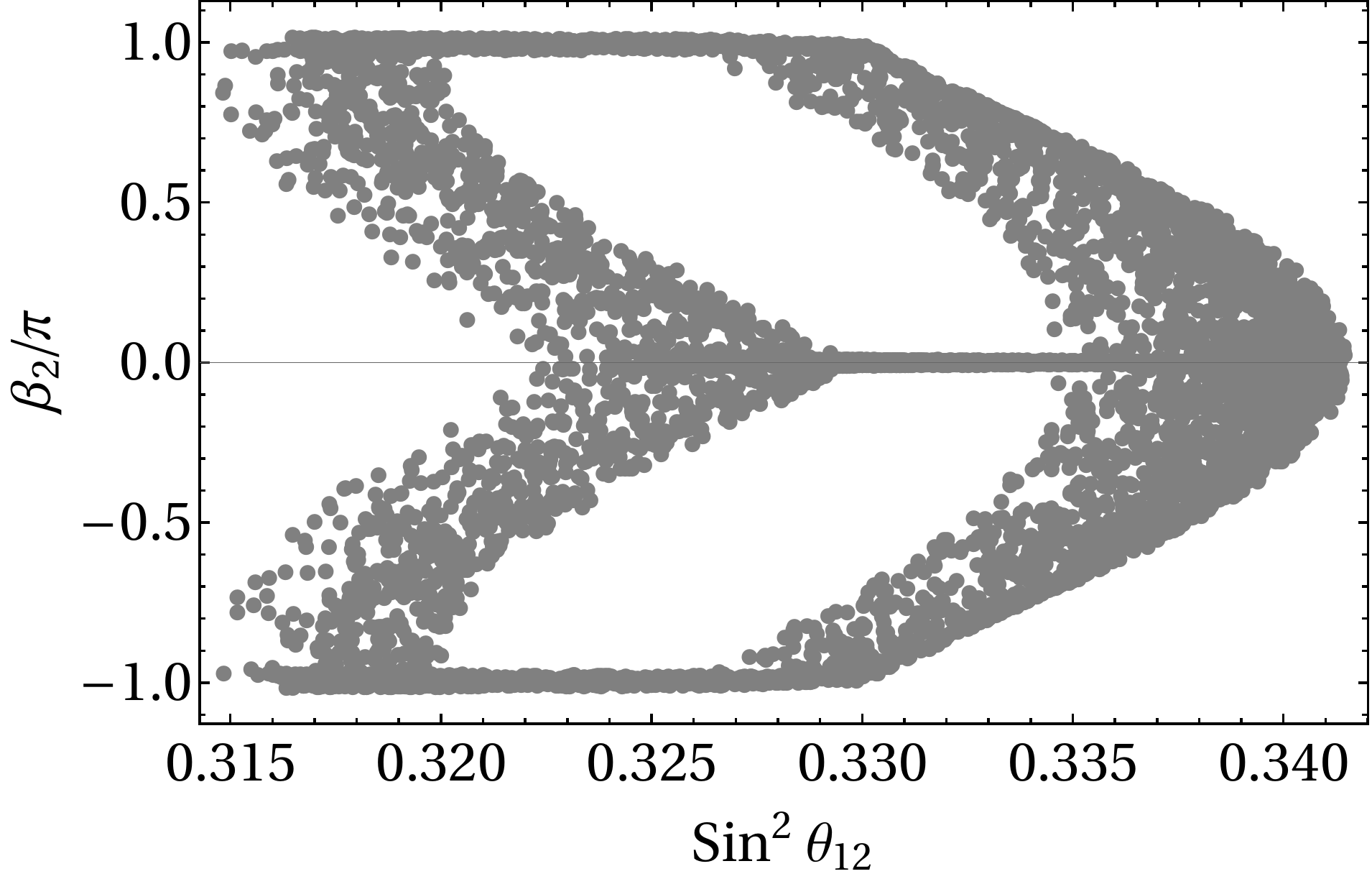}
\caption{Same description as in Fig. \ref{fig:12-13-libres} for $\sigma' = 0$.}\label{fig:23-13-cerolibre}
\end{figure}

In Fig. \ref{fig:23-13-librecero} we show the relation between the $CP$ phases and $\sin^2 \theta_{12}$ for the case $\sigma = 0$. We can observe that $\beta_1$ remains restricted to almost null values in the full range of $\sin^2 \theta_{12}$, while  $\delta_{CP}$ and $\beta_2$ are restricted to small values only for the lower range of $\sin^2 \theta_{12}$. For values of  $\sin^2 \theta_{12} \gtrsim 0.336$,~ $\delta_{CP}$ and $\beta_2$ are almost undetermined, which make this scenario difficult to test.
   
In the case $\sigma' = 0$, we show the plots of the allowed $CP$ phases in terms of the solar angle in Fig. \ref{fig:23-13-cerolibre}. Although $\beta_1$ remains small in this case, values for $\delta_{CP}$ and $\beta_2$ different from zero are allowed in almost the full range of $\sin^2 \theta_{12}$.

\begin{figure}\centering
\includegraphics[scale=0.6]{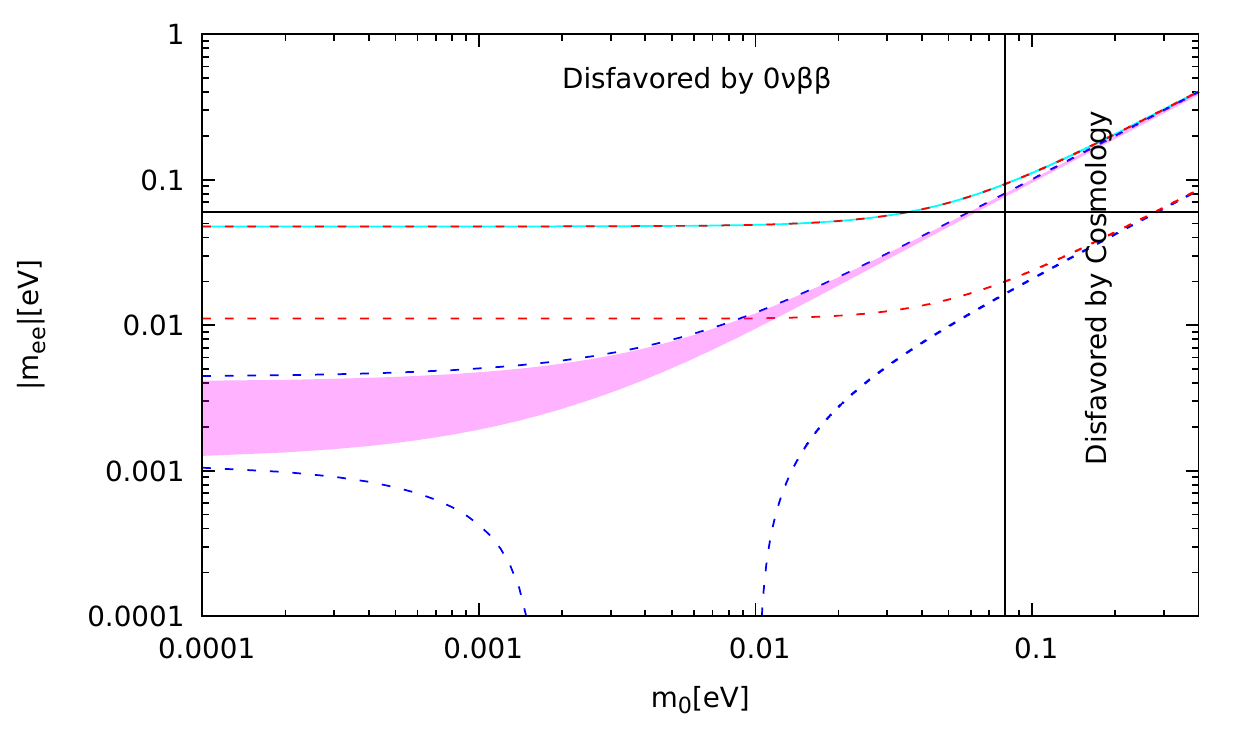} 
\includegraphics[scale=0.6]{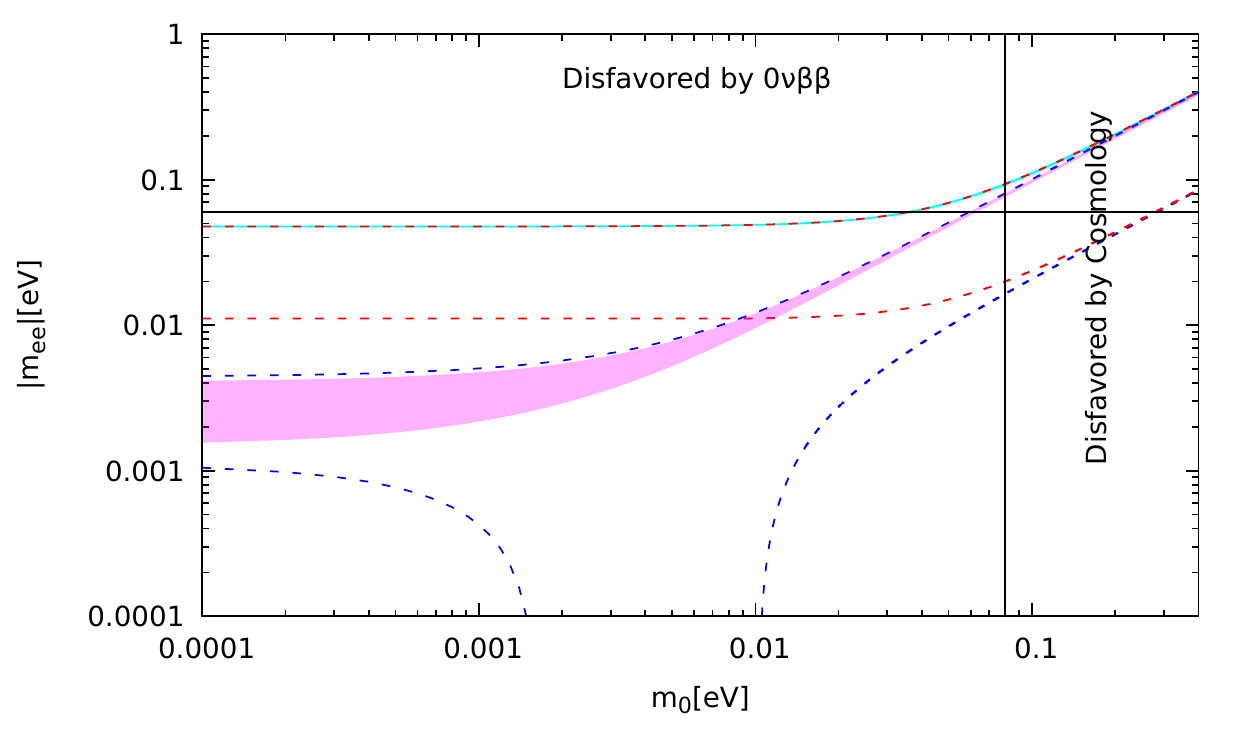} 
\caption{Same description as in Fig. \ref{fig:12-13-0bb} but for $U_{23}(\phi,\sigma)U_{13}(\phi',\sigma')$ (left), and $\sigma^{(\prime)} = 0$ (right).}\label{fig:23-13-0bb}
\end{figure}

 We show in Fig. \ref{fig:23-13-0bb} the predicted regions of $m_{ee}$ for the three cases of interest. In these cases, smallness of $\beta_1$ seems to dominate the behaviour of $m_{ee}$, which supports the similarities between both plots in Fig. \ref{fig:23-13-0bb}. Only a slightly reduction is observed in the region of the NH for $\sigma = 0$ and $\sigma' = 0$, mainly due to the reduction in the allowed values of $\beta_1$ caused by these selections. No visible differences were observed for these cases such that only one plot is presented for both combinations.

%%%%%%%%%%%%%%%%%%%%%%%%%%%%%%%%%%%%%%%%%%%%%%%%%%%%%%%%%%%%%%%%%%%%%%%%%%%%%%%%%%%%%%%%%%%%%%%%%%%%%%%%%%%%%%%%%%%%%%%%%%%%%%%
%%%%%%%%%%%%%%%%%%%%%%%%%%%%%%%%%%%%%%%%%%%%%%%%%%%%%%%%%%%%%%%%%%%%%%%%%%%%%%%%%
\section{Summary}\label{sec:conclusions}

 Although an exact Tri-Bi-Maximal pattern is not consistent with current determinations of neutrino mixing parameters, some deviations can be implemented to restore their compatibility. We have shown that two unitary rotations can be considered as correction matrices to the TBM pattern to obtain a mixing matrix consistent with the global fits, and in some cases, to predict specific regions for Dirac and Majorana $CP$ phases. 
 
 Corrected TBM matrices of the form $U_{TBM} U_{ij}(\phi,\sigma) U_{lm}(\phi',\sigma')$ were analysed. The four introduced parameters were varied to match the neutrino mixings up to $3\sigma$. These parameters were then related to the $CP$ invariants of the neutrino mixing matrix, which allowed to restrict the $CP$ phases to specific regions in terms of one mixing angle. We found that the six combinations $U_{12}U_{13}$, $U_{12}U_{23}$, $U_{13}U_{12}$, $U_{13}U_{23}$, $U_{23}U_{12}$, and $U_{23}U_{13}$ present at least one particular case where the three $CP$ phases could be determined \textit{via} one experimental angle. Future improvements in the determinations of the mixing angles, and an independent determination of the Dirac $CP$ phase, could help to discriminate some of these scenarios.  

 Plots for $|m_{ee}|$ were presented for the various combinations, some of them showed narrow regions in both neutrino mass hierarchies. Some predictions for the IH are very sharp and near the reach of $0\nu\beta\beta$ experiments, allowing the possibility of being readily ruled out with forthcoming results.
    
 Concerning the $\mu-\tau$ symmetry in the mass matrix, we found that only the cases $U_{12}U_{13}$ and $U_{13}U_{12}$ were compatible with the requirement of small deviations from the $\mu-\tau$ symmetry. Despite this restriction helped reducing the predicted regions of the $CP$ phases, the $|m_{ee}|$ values were restricted to the quasi-degenerate hierarchy, which could be discriminated by future bounds on $m_0$ coming from Cosmology and $0\nu\beta\beta$ experiments.
 
We have shown that the so long investigated TBM pattern can be consistent with the experimental determinations of neutrino mixing angles if some deviations are implemented. In addition, some information about $CP$ Majorana phases could be obtained in an indirect way from the restrictions on these correction parameters. Future improvements in neutrino experiments may help to discriminate these scenarios and shed light on the underlying flavor structure of the neutrino sector.   
   
\section*{Acknowledgements}

S.L.T. is grateful to Norberto Granda and Hern\'an Ocampo for the hospitality at the Physics Department of Universidad del Valle.

%\begin{thebibliography}{10}
\bibliographystyle{utphys}
%\bibliography{Bibliography.bib}
%%%%%%%%%%%%%%%%%%%%%%%%%%%%%%%%%%%%%%%%%%%%%%%%%%%%%%%%%%%%%%%%%%%%%%%%%%%%%%
%%%%%%%%%%%%%%%%%%%%%%%%%%%%%%%%%%%%%%%%%%%%%%%%%%%%%%%%%%%%%%%%%%%%%%%%%%%%%%%%%%%%%%%%%%%%%%

\providecommand{\href}[2]{#2}\begingroup\raggedright\endgroup

%%%%%%%%%%%%%%%%%%%%%%%%%%%%%%%%%%%%%%%%%%%%%%%%%%%%%%%%%%%%%%%%%%%%%%%%%%%%%%%%%%%%%%%%%%%

\end{document}